\documentclass[12pt]{article}
\usepackage{epsfig}
\usepackage{comment}
\usepackage{latexsym}
\usepackage{color}
\usepackage{amsmath}
\usepackage{hyperref}

\newcommand{\mysquare}[0]{\raise-.2ex\hbox{{\Large$\Box$}}}
\def\lsim{\mathrel{\rlap {\raise.5ex\hbox{$ < $}}
{\lower.5ex\hbox{$\sim$}}}}
\def\gsim{\mathrel{\rlap {\raise.5ex\hbox{$ > $}}
{\lower.5ex\hbox{$\sim$}}}} \topmargin -1.5cm \textheight=22.5cm \textwidth=16.5cm
\catcode`\@=11
\newcount\hour
\newcount\minute
\newtoks\amorpm
\hour=\time\divide\hour by60 \minute=\time{\multiply\hour by60 \global\advance\minute by-\hour}
\edef\standardtime{{\ifnum\hour<12 \global\amorpm={am}%
        \else\global\amorpm={pm}\advance\hour by-12 \fi
        \ifnum\hour=0 \hour=12 \fi
        \number\hour:\ifnum\minute<10 0\fi\number\minute\the\amorpm}}
\edef\militarytime{\number\hour:\ifnum\minute<10 0\fi\number\minute}
\def\draftlabel#1{{\@bsphack\if@filesw {\let\thepage\relax
   \xdef\@gtempa{\write\@auxout{\string
      \newlabel{#1}{{\@currentlabel}{\thepage}}}}}\@gtempa
   \if@nobreak \ifvmode\nobreak\fi\fi\fi\@esphack}
        \gdef\@eqnlabel{#1}}
\def\@eqnlabel{}
\def\@vacuum{}
\def\draftmarginnote#1{\marginpar{\raggedright\scriptsize\tt#1}}
\def\draft{\oddsidemargin -.2truein
        \def\@oddfoot{\sl preliminary draft \hfil
        \rm\thepage\hfil\sl\today\quad\militarytime}
        \let\@evenfoot\@oddfoot \overfullrule 3pt
        \let\label=\draftlabel
        \let\marginnote=\draftmarginnote
   \def\@eqnnum{(\theequation)\rlap{\k

 ern\marginparsep\tt\@eqnlabel}%
\global\let\@eqnlabel\@vacuum}  }

\newcommand{\be}[0]{\begin{equation}}
\newcommand{\ee}[0]{\end{equation}}
\newcommand{\ba}[0]{\begin{eqnarray}}
\newcommand{\ea}[0]{\end{eqnarray}}

\def\bs{\begin{subequations}}
\def\es{\end{subequations}}

\def\thebibliography#1{%
\vskip 0.5cm \centerline{\bf \Large References}
\list{%
[\arabic{enumi}]}{\settowidth\labelwidth{[#1]} \leftmargin\labelwidth \advance\leftmargin\labelsep
\usecounter{enumi}}
\def\newblock{\hskip .11em plus .33em minus .07em}
\sloppy\clubpenalty4000\widowpenalty4000 \sfcode`\.=1000\relax}

\renewcommand{\theequation}{\arabic{section}.\arabic{equation}}

\renewcommand{\section}{\setcounter{equation}{0}\@startsection
{section}{1}{0mm}{-\baselineskip}{0.5\baselineskip} {\normalfont\Large\bfseries}}

\renewcommand{\subsection}{\@startsection
{subsection}{2}{0mm}{-\baselineskip}{0.5\baselineskip} {\normalfont\large\bfseries}}

\renewcommand{\subsubsection}{\@startsection
{subsubsection}{3}{0mm}{-\baselineskip}{0.5\baselineskip} {\normalfont\normalsize\slshape}}

\usepackage{amssymb,amsfonts}
\usepackage{graphicx}
\usepackage{cite}

\newcommand{\red}{}

\newcommand{\Co}{\mathbb{C}}
\newcommand{\Pb}{\mathbb{P}}
\newcommand{\Z}{\mathbb{Z}}
\newcommand{\Ka}{K{\"a}hler }

\newcommand{\sign}{{\rm sign}}
\newcommand{\abs}{|}

\newcommand{\ie}{{\em i.e. }}
\newcommand{\where}{\mbox{where}}
\newcommand{\with}{\mbox{with}}
\newcommand{\when}{\mbox{when}}
\renewcommand{\and}{\mbox{and}}

\newcommand{\N}{{\cal N}}
\newcommand{\M}{{\cal M}}
\newcommand{\A}{{\cal A}}
\newcommand{\B}{{\cal B}}
\newcommand{\C}{{\cal C}}
\newcommand{\K}{{\cal K}}
\newcommand{\Zc}{{\cal Z}}
\newcommand{\D}{{\cal D}}
\newcommand{\E}{{\cal E}}
\renewcommand{\o}{\overset{\circ}}
\newcommand{\oo}{\overset{\circ\circ}}
\newcommand{\dil}{\phi_{\mbox{\tiny dil}}}
\newcommand{\tg}{\tilde g}
\newcommand{\tk}{\tilde k}
\newcommand{\tm}{\tilde m}
\newcommand{\ve}{\varepsilon}

\newcommand{\bea}{\begin{eqnarray}}
\newcommand{\eea}{\end{eqnarray}}
\newcommand{\dis}{\displaystyle}

\def\cO{{\cal O}}
\def\no{\nonumber}
\def\half{ {1\over 2}}
\def\ab{[{}^a_b]}

\def\abb{[{}^{\bar a}_{\bar b}]}

\def\p{\partial}

\begin{document}
\begin{titlepage}
\begin{flushright}
CPHT--RR079.0709,
LPTENS--09/24,
August 2009
\end{flushright}

\vspace{1mm}

\begin{centering}
{\bf\huge Cosmological Phases of the String Thermal Effective Potential$^\ast$}\\

\vspace{6mm}
 {\Large F. Bourliot$^{1}$, J. Estes$^{1}$, C. Kounnas$^{2}$ and H. Partouche$^1$}

\vspace{2mm}

$^1$  {Centre de Physique Th\'eorique, Ecole Polytechnique,$^\dag$
\\
F--91128 Palaiseau cedex, France\\
{\em Francois.Bourliot@cpht.polytechnique.fr} \\
{\em John.Estes@cpht.polytechnique.fr}\\
{\em Herve.Partouche@cpht.polytechnique.fr}}

\vspace{2mm}

$^2$ Laboratoire de Physique Th\'eorique,
Ecole Normale Sup\'erieure,$^\ddag$ \\
24 rue Lhomond, F--75231 Paris cedex 05, France\\
{\em  Costas.Kounnas@lpt.ens.fr}

\end{centering}

$~$\\
\centerline{\bf\Large Abstract}
\vskip .1cm
\noindent
In a superstring framework, the free energy density ${\cal F}$ can be determined unambiguously \red {at the full string level}
once supersymmetry is spontaneously broken via geometrical fluxes. We show explicitly that
only the moduli associated to the supersymmetry breaking may give relevant contributions. All other
spectator moduli $\mu_I$ give exponentially suppressed contributions for relatively \red{\emph{small} (as compared to the string scale)} temperature $T$ and supersymmetry breaking scale $M$.  More concisely, for $\mu_I > T$ and $M$, ${\cal F}$ takes the form
$$
{\cal F}(T,M; \mu_I)={\cal F}(T,M)+{\cal O}\left[~{\rm exp}\left(- {\mu_I\over T}\right),~{\rm exp}\left(- {\mu_I\over M}\right)~\right]\,.
$$
\red{We study the cosmological regime where $T$ and $M$ are} below the Hagedorn temperature scale $T_H$.  In this regime, ${\cal F}$ remains finite for any values of the spectator moduli $\mu_I$. We investigate extensively the case of one spectator modulus
\red{$\mu_d$ corresponding to $R_d$,} the radius-modulus field of an internal compactified dimension. We show that its thermal effective potential $V(T,M; \mu)={\cal F}(T,M; \mu)$ admits five  phases, each of which can be described by a distinct \red{but different} effective field theory. For late cosmological times, the Universe is attracted
to a ``Radiation-like evolution'' with $M(t)\proptoÊT(t)\propto 1/a(t)\propto t^{-2/d}$. The spectator modulus $\mu(t)$ is stabilized either to the stringy enhanced symmetry point where \red{$R_d=1$}, or fixed at an arbitrary constant $\mu_0>T,M$.
For arbitrary boundary conditions at \red{some initial time,} $t_E$, $\mu(t)$ may pass through more than one effective field theory phase before its final attraction.

\vspace{3pt} \vfill \hrule width 6.7cm \vskip.1mm{\small \small \small
  \noindent $^\ast$\ Research
partially supported by the ERC Advanced Grant 226371, ANR  contract 05-BLAN-0079-02, CNRS PICS contracts  3747 and 4172, and the Groupement d'Int\'er\^et Scientifique P2I.\\
$^\dag$\ Unit{\'e} mixte du CNRS et de l'Ecole Polytechnique,
UMR 7644.}\\
 $^\ddag$\ Unit{\'e} mixte  du CNRS et de l'Ecole Normale Sup{\'e}rieure associ\'ee \`a
l'Universit\'e Pierre et Marie Curie (Paris 6), UMR 8549.

\end{titlepage}
\newpage
\setcounter{footnote}{0}
\renewcommand{\thefootnote}{\arabic{footnote}}
 \setlength{\baselineskip}{.7cm} \setlength{\parskip}{.2cm}

\setcounter{section}{0}


\section{Introduction}
String theory provides a framework to obtain a sensible theoretical
description of the cosmological evolution of our Universe. Nowadays,
it is the only known framework in which the quantum
gravity effects are under control \cite{GSW}, at least for certain physically relevant cases.
Following the stringy cosmological approach developed recently in Refs \cite{Cosmo-1,Cosmo-2, Francois},
the classical  string vacuum is taken to be supersymmetric with a fixed amount of
supersymmetries defined in flat space-time.

This initial choice does not give rise to any cosmological evolution. In the presence of supersymmetry, the quantum corrections to the gravitational background would lead to a flat space-time, or would modify it at most to Anti-de Sitter, domain walls or gravitational wave  backgrounds respecting a time-like or light-like killing symmetry.
The above cosmological obstructions are however physically irrelevant for two fundamental reasons:

$\bullet$ Firstly, supersymmetry is broken in the real world, (at least spontaneously and not explicitly), at a characteristic supersymmetry breaking scale $M$.

$\bullet$ Secondly, in the case of thermal cosmologies,
the supersymmetry is effectively (spontaneously) broken at the temperature scale $T$.

Both the $M$ and $T$ supersymmetry breaking scales induce at the quantum level a non-trivial free energy density ${\cal F}(T,M)$, which plays the role of an effective thermal  potential $V(T,M)={\cal F}(T,M)$ that modifies the gravitational and field equations, giving rise to non-trivial cosmological solutions, as has been explicitly shown  in  Refs \cite{Cosmo-1,Cosmo-2, Cosmo-0}. Both the supersymmetry breaking and finite temperature phenomena can be implemented in the framework of superstrings \cite{AtickWitten,RostKounnas,AKADK} by introducing non-trivial ``fluxes" in the initially supersymmetric vacua. Furthermore, in the case where supersymmetry is spontaneously broken by ``geometrical fluxes" \cite{GeoFluxes,OpenFluxes},  the free energy ${\cal F}(T,M)$ is under control and is calculable at the \red{\emph{full}} string level, free of any infrared and ultraviolet ambiguities \cite{Cosmo-1, Cosmo-2}.  This is true, provided $T$ and $M$ are below a critical value close to the string mass scale, the so-called Hagedorn temperature $T_H$ \cite{AtickWitten,RostKounnas,AKADK,Hagedorn}.  In the framework of stringy-thermal cosmologies, $T\simeq T_H$ corresponds to very early times when we are facing non-trivial stringy singularities indicating a non-trivial phase transition at high temperatures \cite{AtickWitten,AKADK,GV,BV,GravFluxes} at  time $t_H$.
In the literature, there are many speculative proposals concerning the nature of this
transition \cite{AtickWitten,AKADK,GV,BV,Kaloper:2006xw,GravFluxes,MSDS}.

A way to bypass the Hagedorn transition ambiguities was proposed in Ref. \cite{Francois}. It consists of assuming the emergence of ($d-1$) large space-like directions for times $t\gg t_H$,  describing the ($d-1$)-dimensional space of the Universe, and  possibly some internal space directions of an intermediate size characterizing the scale $M$ of the spontaneous breaking of supersymmetry via geometrical fluxes \cite{GeoFluxes,OpenFluxes}.  Within these assumptions, the  ambiguities of the ``Hagedorn transition exit at $t_E$" can be parameterized, for $t\ge t_E$,  in terms of initial  boundary condition data at  $t_E\ge t_H$. In this way, the intermediate cosmological era $ t_E\le t \le t_W$, \ie after the  ``Hagedorn transition exit"  and before the electroweak symmetry breaking phase transition at $t_W$, was extensively  studied in Ref.  \cite{Francois} in the case of $d=4$.  An output of the present analysis is that the cosmological ``radiation-like" evolution found in Refs \cite{Cosmo-1, Cosmo-2,Cosmo-0,AKradiation} generalizes to a ``Radiation\red{-like} Dominated Solution" (RDS) in $d$-dimensional space-time,
 \be
 \label{RDS}
{\rm RDS}^d~:~~~~~ M(t)\propto(t)\propto 1/a(t)\propto t^{-2/d},~~~~{\rm for} ~~~~t_E\le t\le t_W,
 \ee
and is unique at late times in certain physically relevant  supersymmetry breaking schemes.   \red{As a necessary and sufficient consistency requirement, we note that in this intermediate cosmological regime $T_H \gg T,M \gg T_W$, the smallness of the space-time curvature scales, ${H^2=(\dot a}/a)^2$ and $\dot H$, the dilaton and the evolving radii scales, ${\dot \phi}_{\mbox{\tiny dil}}\!\!\!{}^2$, ${\ddot \phi}_{\mbox{\tiny dil}}$,$({\dot R}_I/R_I)^2$, $\ddot R_I/R_I$,  are guaranteed to be small  ($\le {\cal O}(T^d, M^d)$), thanks to the ``attractor mechanism" towards the RDS in late cosmological times.  In particular, they are all decreasing at late cosmological times and so our quasi-static approximation becomes better and better as time passes.  In addition, our perturbative approximation becomes better and better as time progress due to the falling of the dilaton.}

\red{We point out that the evolution is radiation-like in the sense that the external space-time's evolution is identical to that of a radiation dominated universe.  However, in our case the expansion is driven not only by radiation, but also by the coherent motion of the supersymmetry breaking modulus $M$.  We would also like to stress that a key result of \cite{Francois} was that the $RDS^4$ is actually an attractor of the dynamics.  The important consequence of this attractor, is to wash out the dependence of the cosmological evolution on the choice of initial boundary conditions, which were used to parameterize our ignorance of the physics involved in the Hagedorn transition.  In this paper we show that the attractor naturally extends to the higher-dimensional $RDS^d$.
}

Although this analysis was done in the framework of initial vacua with $\N_4=2$ supersymmetry,
the claim is that it will still be valid in more realistic  models with initial $\N_4=1$ supersymmetry \cite{N=1Estes}. We would like to stress here that the limitation  $ t \le t_W$ in the infrared regime  follows from the appearance in the low energy effective field theory of a new scale, namely the  ``infrared renormalization group invariant transmutation scale $Q$",  at which the supersymmetric standard model Higgs (mass)$^2$ becomes negative, (no-scale radiative breaking of $SU(2)\times U(1)\to U(1)_{\rm em}$ \cite{Noscale, NoscaleTSR}). $Q$ is irrelevant as long as $M,T\gg Q$; however it becomes relevant and stops the $M(t)$ evolution when $T\simeq Q$ at $t\simeq t_W$ \ie when the electroweak breaking phase transition takes place.  Although the physics for $t\gg t_W$ is of main importance in particle physics and in inflationary cosmology at $t_W$, it will not be examined in this work. The main reason for us is its strong dependence on the initial vacuum data which screens interesting universality properties.  \red{\emph {We therefore work
in the intermediate cosmological era $t_E\le t \le t_W$ or $T_H\gg T\gg Q$, \ie after the Hagedorn phase transition and before the electroweak one.}} In this regime the transmutation scale $Q$ can be consistently neglected and, furthermore, the Hagedorn transition ambiguities are taken into account in terms of initial boundary conditions (IBC) after the ``Hagedorn transition exit''  at $t_E$.  \red{This scenario gives a dynamical explanation of the smallness of the supersymmetry breaking scale as compared to the string or Planck scales.  Indeed, extrapolating the $RDS^{d=4}$ up to the low energy regime where $T={\rm \cal O}(1~TeV)$  one finds, (thanks to the attractor mechanism),  that the natural value of the supersymmetry breaking scale $M(t)$ is naturally small and around the electroweak phase transition, independently of its initial value at early cosmological times.}

\red {The only known supersymmetry breaking mechanism that can be unambiguously adapted at the string perturbative level, is the one we consider here where supersymmetry is spontaneously broken via ``geometrical fluxes".  This choice implies the existence of at least one relatively large compact dimension (the one which is associated to supersymmetry breaking). This is by far not in contradiction with experimental results both in particle physics and cosmology. For several years, the possibility of ``large extra dimensions'' has attracted the attention of the particle physics community; the future data analysis at the LHC and elsewhere includes searches for signals indicating the existence of large extra dimensions at scales $1/R\sim {\rm \cal O}(1~TeV)$, which is the characteristic prediction of supersymmetry breaking via geometrical fluxes. }

\red{Many other choices for supersymmetry breaking exist.  However in most cases it is not well known, in our days, the precise stringy corrections to all orders in $\alpha'$ that are necessary in order to study the intermediate cosmological regime. In all other supersymmetry breaking mechanisms we are forced to work in the effective supergravity framework.   Our results may help to make more general stringy approaches possible in the future.  There are indications that our results can be converted to other string vacua, utilizing string/string and M-theory dualities.  In particular the geometrical fluxes are mapped to other types of fluxes, for instance the three form R-R and NS-NS fluxes in type IIB - orientifolds. In this respect, the exact stringy approach in the intermediate cosmological regime gives us profound (non-perturbative) information via M-theory and string dualities.}

In addition to generalizing the results of \cite{Francois} to arbitrary dimension, we analyze the time behavior of {\it the spectator moduli  not participating in the breaking of supersymmetry}.  Following Refs \cite{Cosmo-1, Cosmo-2,Cosmo-0}, one can show that only the supersymmetry breaking moduli $M(t)$ and $T(t)$ can give a relevant contribution to the free energy density ${\cal F}$.  Intuitively, all other moduli  $\mu_I$ are either attracted and stabilized to the ``stringy" extended gauge symmetry points, close to the string scale $\mu_I\sim M_{\rm string}$,  or are effectively frozen to an arbitrary value such that $\mu_I \gg T$ and $M$,  giving rise to exponentially suppressed contributions:
\be
{\cal F}(T,M; \mu_I)={\cal F}(T,M)+{\cal O}\left[~{\rm exp}\left(- {\mu_I\over T}\right),~{\rm exp}\left(- {\mu_I\over M}\right)~\right].
\ee
One point of the present paper is to explicitly verify this intuition for the moduli coming from the spectator tori.  Indeed, the supersymmetry breaking moduli generate a non-trivial potential for the spectator moduli and freeze them as expected.
Considering the effect on a single spectator modulus ($\mu \propto1/R_d$) in the case of the heterotic string, the thermal effective potential $V(T,M; \mu)$ admits five distinct phases, each of which can be described by a \red{different} effective field theory.  \red{The interesting result is that by using a string theory framework and working at the \emph{full} string level, we are able to link together, within a single framework, the different effective field theories.  Furthermore, the main result of this paper is to derive at the string perturbative level, (however exact in $\alpha'$), the full string free energy ${\cal F}(T,M; \mu_I)$ as a functional of the supersymmetry breaking moduli $T,M$, the string coupling constant modulus $\dil$ and ``spectator moduli" $\mu_I$, for a certain class of string vacua where the spontaneous breaking of supersymmetry is induced by geometrical fluxes.}

The form of the potential is sketched in Fig. \ref{fig_potential} and the phases are summarized as follows ($T\propto {1/R_0},~M\propto {1/R_9}$):
%
\begin{figure}[h!]
\begin{center}
\vspace{0.3cm}
\includegraphics[height=5.7cm]{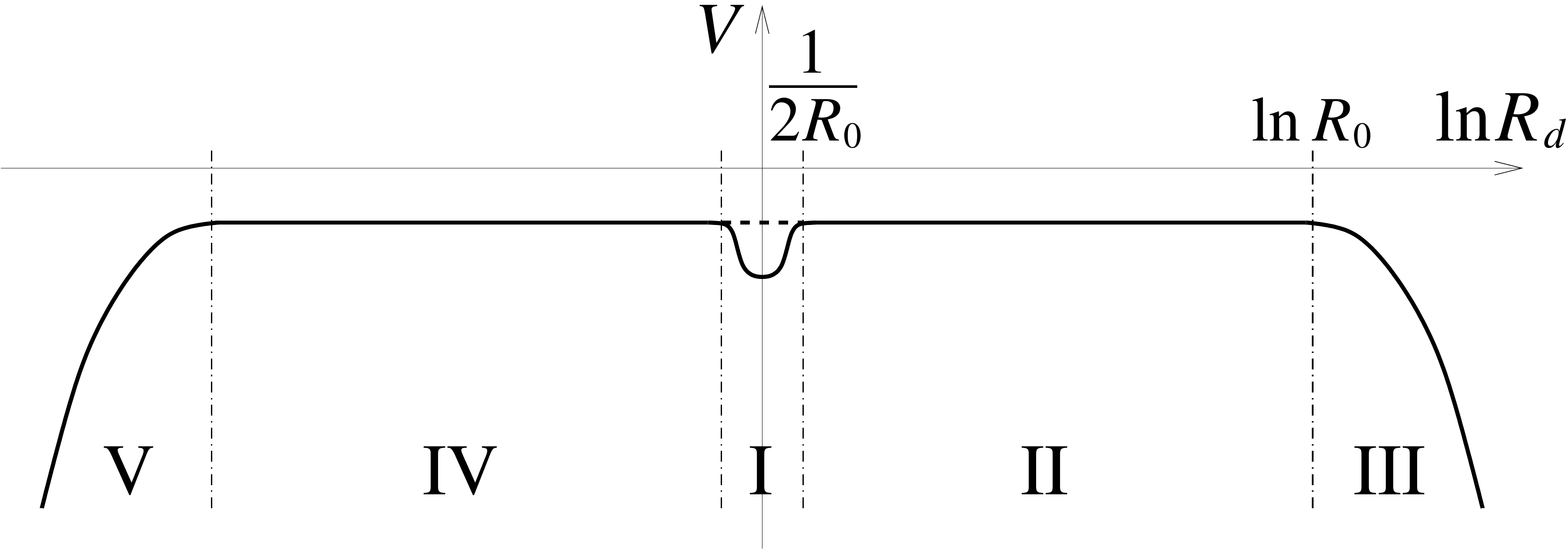}
\caption{\footnotesize \em Qualitative shape of the effective potential
$V$ versus
$\ln R_d $, with $T$, $M$ (and the dilaton in Einstein frame) fixed.  When
$\ln R_d$ varies, five distinct phases can arise in the heterotic case.
i) The Higgs phase {\rm I}, ii) the flat potential phases {\rm II}, {\rm
IV} and iii) the higher-dimensional phases {\rm III}, {\rm V}.
The phases {\rm IV} and {\rm V} are T-dual to {\rm II} and {\rm III}.
 In the type II case, the phase {\rm I} does not exist and the plateaux
{\rm II} and {\rm IV} are connected.  The plot is valid for $T>M$, for $T<M$ one simply replaces $R_0$ with $R_9$.}
\label{fig_potential}
\end{center}
\vspace{-0.5cm}
\end{figure}
%

\indent \indent {\it \bf  I. Higgs phase:} With $\dis \left\abs R_d-{1\over R_d}\right\abs < {1\over R_0} ~{\rm and/or}~{1\over R_9}.$

\noindent This phase contains the stringy extended symmetry point at the self-dual point $R_d=1$. The appropriate effective field theory description of this phase is  in terms of a $d$-dimensional theory of gravity coupled to an $SU(2)$ gauge theory. $R_d$ is dynamically stabilized at this point.  Such a notion of moduli stabilization, has been studied in the literature before \cite{Watson:2004aq,Patil:2004zp,BV}. Here we demonstrate  such moduli stabilization in the context of the heterotic superstring by an explicit  computation of the effective potential.

\indent \indent {\it \bf II. Flat potential phase:} With $\dis {1\over R_0}~ {\rm and} ~{1\over R_9}< R_d - {1\over R_d}  < R_0~ {\rm and} ~R_9.$

\noindent Here, the appropriate  effective field theory description is in terms of a $d$-dimensional theory of gravity coupled to an $U(1)$ gauge field.

\indent\indent {\it \bf III. Higher-dimensional phase:} With $R_0~ {\rm and/or} ~R_9  <R_d.$

\noindent For macroscopic values of the spectator modulus $R_d$, the appropriate effective field theory description is the $(d+1)$-dimensional theory of gravity.  The modulus $R_d$ becomes the $\hat g_{dd}$ component of the metric, $\hat g_{dd}=(2\pi R_d)^2$ and the evolution is attracted to that of an RDS in $d+1$ dimensions.

One may also consider the case in which the radius $R_d$ is still internal.  For large enough values of $R_d$ so that we can neglect terms of order $(R_0 / R_d)^d$ and $(R_9 / R_d)^d$, the evolution is attracted to an RDS$^{d+1}$ for a long period of time.  However, at late times $R_0$ and $R_9$ always catch $R_d$ and the solution is ultimately attracted to the RDS$^d$ of phase II.

\indent\indent{\it \bf IV. Dual flat potential phase:} With $\dis {1\over R_0}~ {\rm and} ~{1\over R_9}<  {1\over R_d} -R_d  < R_0~ {\rm and} ~R_9.$

 \noindent The effective theory description is T-dual to that of phase II. The light degrees of freedom are the winding modes  instead of the Kaluza-Klein momenta of phase II.

\indent\indent {\it \bf V.  Dual higher-dimensional phase:} With $\dis  R_0~ {\rm and/or} ~R_9<{1\over R_d}. $

\noindent This phase is T-dual to phase III. Its properties are derived from the ones of phase III under the replacement $R_d\to 1/R_d$ and phase II $\to$ IV. In particular, the dual effective field theory is a $(d+1)$-dimensional theory of gravity, with $\hat g_{dd}=(2\pi/R_d)^2$.

 The above different phases of a common string setting  {\it cannot be described in the context of a single field theory}. This is due to the necessary presence of the string winding modes.  In a field theory framework, only the phases II and III (or IV and V) can be described by a common field theory.
The winding modes are particularly important for the stabilization of the modulus in phase I at the extended symmetry point, and furthermore for the description of the T-dual phases IV and V.  In contrast to field theory, string theory naturally interpolates between these various phases, due to the generation of an effective potential in the presence of temperature and spontaneous supersymmetry breaking.

For each phase, there exists an RDS as in (\ref{RDS}).  We show that these solutions are stable against small perturbations and that for arbitrary IBC close to an RDS, the cosmological evolution is attracted to this RDS.  In \cite{Francois}, the spectator moduli were taken to be frozen and it was shown that under this hypothesis the RDS is a global attractor.  Taking into account the existence of gravitational friction for an expanding universe, we expect in the present work with dynamical spectator moduli such as $R_d$, the results of \cite{Francois} to generalize so that the  evolution is always attracted to the RDS of one of the five phases.

In the case of type II strings, the phase I does not exist perturbatively, so that the phases II and IV combine into a single plateau.  This is due to the lack of massless states necessary to enhance the $U(1)$ to $SU(2)$.  However, by heterotic-type II duality, we expect such an enhancement to exist non-perturbatively, so that  all five phases should exist at the non-perturbative level.  The effects correspond to the addition of branes whose separation is governed by the spectator moduli.

The organization of the paper is as follows.  In section 2, we discuss in more details the specific setup analyzed in this paper.  The thermal effective potential is given and it is shown to have the five phases discussed above.  In section 3, we show that an RDS solution exists in each  phase.  In addition, we show the stability of the solutions against small perturbations.  In section 4, we briefly discuss the role of non-perturbative objects in the type II string theory.  In section 5, we summarize our results and discuss further avenues of research. In appendix A, the thermal partition functions for the heterotic and type II strings are presented, together with their asymptotic properties to be used in the different phases. The gravity and field equations are given for each effective field theory phase in appendix B.


\section{Effective thermal potential in superstrings with\\spontaneously broken supersymmetry}
\label{potential}

In the presence of temperature, the one-loop partition function of both the heterotic and type II superstrings is non-vanishing and yields the one-loop effective potential at finite temperature.  In addition, spontaneous supersymmetry breaking is induced by the presence of geometrical fluxes\cite{Cosmo-1, Cosmo-2} along the internal cycles of the background manifold.  We introduce these fluxes via the generalization to the context of string theory of Scherk-Schwarz \cite{SS} compactifications in field theory\cite{Rohm,KouPor}. They induce further contributions to the one-loop partition function, which persist even at zero temperature\cite{Cosmo-1, Cosmo-2}.  Due to various ways of introducing the fluxes, there are multiple supersymmetry breaking configurations for the same initially supersymmetric background.  In \cite{Cosmo-1, Cosmo-2}, such one-loop thermal effective potentials were derived in the limit of small temperature $T$ and small supersymmetry breaking scale $M$ for the heterotic and type II superstrings  compactified on $T^6$  and  $T^2 \times \dis {T^4 \over  \Z_2}$ orbifolds.  The partition functions were calculated for small but otherwise arbitrary temperature and supersymmetry breaking scale, while the remaining moduli were taken to be frozen close to the string scale.

We want to  relax the latter hypothesis and examine the behavior of the spectator moduli in the presence of temperature and supersymmetry breaking.  In appendix A, we compute  the partition functions for the heterotic (\ref{part2het}) and type II  (\ref{part2ii}) cases.  The background manifold is of the form $S_E^1 \times T^D \times T^n$ (or  $S_E^1 \times T^D \times \dis {T^4 \over  \Z_2} \times T^n$ in the orbifold models), where $S_E^1$ is the compact Euclidean time circle and the $T^n$ torus involves the geometrical fluxes which generate the breaking of supersymmetry. The $T^D$ spectator moduli  are not participating in the breaking of supersymmetry.  We take both the temperature and supersymmetry breaking scales to be small, while allowing the $T^D$ spectator moduli to remain arbitrary.  This enables us to study the resulting effects of the effective thermal potential  on the $T^D$ moduli.

For simplicity, we specialize to the following 10-dimensional Euclidean background  that contains:
\begin{itemize}
\item The Euclidean time direction, with radius $R_0$ which determines the temperature $T$.
\item The $1,...,d-1$ directions, which are taken to be very large and form, together with the time, a $d$-dimensional space-time.
\item The circle $S^1(R_d)$, with  arbitrary radius. For small $R_d$, $S^1(R_d)$ is considered as part of the internal compactified space. For macroscopic $R_d$, however,  $S^1(R_d)$ becomes part of a space-time of dimension $d+1$. By macroscopic we mean that we can probe it with current experiments.  $R_d$ is the only ``spectator'' radius whose dynamics is taken into account.
\item The $n=1$ circle  involved  in the spontaneous breaking of supersymmetry. We take it to be along the compact direction $9$, with radius $R_9$.
\item The remaining compact directions, with radii $R_{d+1},...,R_8$. They are taken to be fixed close to the string scale. (In the orbifold models, the $\dis {T^4\over \Z_2}$ factor spans the directions $5,\dots, 8$.  Its dynamics are consider in the companion paper \cite{N=1Estes}.)
\end{itemize}
Utilizing the general expressions associated with the heterotic and type II partition functions given in  appendix A, we can easily obtain the ones associated to the background chosen above, namely
\be
\label{background}
S^1(R_0)\times T^{d-1}\times S^1(R_d)\times \M \times S^1(R_9) ,
\ee
where $\M=T^{8-d}$ or $\dis T^{4-d}\times {T^4\over \Z_2}$. The heterotic (type II) models admit a supersymmetry characterized by 16 or 8 (32 or 16) supercharges, which are spontaneously broken by the ``stringy Scherk-Schwarz compactifications'' in the directions 9 and 0\cite{Cosmo-1, Cosmo-2}. The scales of supersymmetry breaking $M$ and temperature $T$ are characterized by $1/R_9$ and $1/R_0$, respectively.  We take $R_0$ and $R_9$ to be large, but still much smaller than the radii of the $T^{d-1}$ torus so that we have the following inequality
\bea
R_1,...,R_{d-1} \gg R_0,R_9 \gg 1.
\eea

As long as $R_d$ is smaller than the size of the external space $T^{d-1}$, it is more convenient  to express the effective field theory action $S$ in terms of fields, which have a natural interpretation in $d$ dimensions.  We are interested in isotropic and homogeneous backgrounds. More specifically, we take the gauge fields to be pure gauge and the remaining scalar fields and the space-time metric to depend only on time.  \red{This will have the advantage that after such a reduction, the different effective field theories will be describable within a single framework.}  The backgrounds we will consider are non-trivial for the $d$-dimensional metric $g_{\mu\nu}$, the $d$-dimensional dilaton $\dil$, and the moduli fields. However, since we allow $R_d$ to vary arbitrarily in size, it may become of the order of the external $T^{d-1}$ radii, so that $S^1(R_d)$ should  be considered as a part of a $(d+1)$-dimensional space-time.  In this case the effective action $S$ is naturally expressed in terms of redefined fields and space-time metric in $d+1$ dimensions.

In appendix \ref{sececmo}, the dimensional reduction from 10 dimensions to $d$ dimensions is carried out explicitly and the resulting action in Einstein frame is given in (\ref{Aaction}).  The case we are considering here has $n=1$, $\Delta = 9-A-d$ (where $A=0$ in the toroidal models and $A=4$ in the orbifold ones) and ${\cal D} = d$. The resulting action is
\be
\label{Scompact}
S=\int d^dx\sqrt{-g}\left( {R\over 2}-{1\over 2}(\partial \Phi)^2-{1\over 2}(\partial \phi_\bot)^2-{1\over 2}(\partial \zeta)^2 +P\right)\, ,
\ee
where we have defined the normalized fields,
\be
\label{Phiphibot}
\Phi:={2\over \sqrt{(d-2)(d-1)}}\, \dil-\sqrt{d-2\over d-1}\, \eta\; , \qquad \phi_\bot:={2\over \sqrt{d-1}}\, \dil +{1\over \sqrt{d-1}}\, \eta
\ee
along with
\be
\label{scalars}
\zeta:=\ln R_d\; , \qquad \eta:=\ln R_9.
\ee
The remaining moduli are taken to be fixed close to the string scale.

The source $P$ is a pressure equal to $-\cal F$, the free energy density (see Eq. (\ref{p=-F})). It is the opposite of the one-loop effective potential at finite temperature and is related to the one-loop partition function as,
\be
P=e^{{{2d\over d-2}\dil}}\,  {Z\over V_{0,...,d-1}} ,
\ee
where $V_{0,...,d-1}$ is the $d$-dimensional Euclidean volume (in string frame) and $Z$ is the partition function computed in appendix A.
In the next sub-section we shall give the exact form of $P$ in terms of a convenient set of variables.

\subsection{Specific form of the effective thermal potential}
\label{form1}

For the heterotic case, the partition function is given in Eq. (\ref{part2het}) which can be re-written in the following  convenient form
\be
\label{Z}
Z=\left(\prod_{i=0}^{d-1}R_i\right) R_9{2^4\over 2}\sum_{\tg_0+\tg_9=1}\left( \Zc_{\mbox{\tiny generic}}+\Zc_{\mbox{\tiny enhanced}}\right) .
\ee
For the type II case, Eq.  (\ref{part2ii}), there is no contribution $\Zc_{\mbox{\tiny enhanced}}$.

In the heterotic case, $\Zc_{\mbox{\tiny enhanced}}$ is generically suppressed except when $R_d$ is close to its self-dual point, where an enhancement of the gauge group $U(1)\to SU(2)$ occurs. The contribution $\Zc_{\mbox{\tiny generic}}$ for generic $R_d$ can be written in two equivalent forms\footnote{In the notations of appendix A, these two forms correspond to $\Delta=9-A-d$ and $\Delta=8-A-d$, where $A=0$ for the toroidal models and $A=4$ for the orbifold ones.}, related to one another by a Poisson resummation of the momentum lattice index $m_d$ of the circle $S^1(R_d)$.

In the Einstein frame, the temperature $T$ and supersymmetry breaking scale $M$ are dressed by the dilaton field  $\dil$  and are  given by Eqs (\ref{TD}) and (\ref{MD}):
\be
\label{TMz}
T={e^{{2\dil\over d-2}}\over 2\pi R_0\, }\; , \qquad M={e^{{2\dil\over d-2}}\over 2\pi R_9\, }\equiv {e^{\sqrt{d-1\over d-2}\Phi}\over 2\pi}\; .
\ee
Observe that in the ratio $\dis{M\over T}={R_0\over R_9}$, the $\dil$ dependence drops out. The expression for
$P$ gets simplified drastically once it is written in terms of the complex structure modulus $z$,
\be
z:=\ln {M\over T}=\ln {R_0\over R_9}.
\ee
In terms of the independent variables $\{T,z,\eta,\zeta\}$,
the pressure $P$ takes the factorized form
\be
\label{pand}
P(T,z,\eta,\zeta) \equiv T^d\, p(z,\eta,\zeta),
\ee
with $\zeta$ and $\eta$ defined in (\ref{scalars}).
Furthermore, $p$ can be written in terms of functions with natural interpretations either in $d$ or $d+1$ dimensions.  (In Eq. (\ref{part2het}), the first case corresponds to  $\Delta =9-A-d$ and the second to $\Delta = 8-A-d$.)
In the heterotic case,  the two equivalent forms for $p$ are:
\be
\label{p}
\begin{array}{ll}
p(z,\eta,\zeta)&\!\!\! =n_T\, \left[\hat f^{(d)}_T(z)+k^{(d)}_T(z,\eta-\abs\zeta\abs)\right]+n_V\, \left[\hat f^{(d)}_V(z)+k^{(d)}_V(z,\eta-\abs\zeta\abs)\right]\\ \\&~~+\tilde n_T\, g^{(d)}_T(z,\eta, \abs\zeta\abs)+\tilde n_V\, g^{(d)}_V(z,\eta, \abs\zeta\abs)\\ \\
&\!\!\!= e^{\abs\zeta\abs-\eta-z}\left[n_T\, f^{(d+1)}_T(z,\eta-\abs\zeta\abs)+ n_V\, f^{(d+1)}_V(z,\eta-\abs\zeta\abs)\right]\\\\
&~~+\tilde n_T\, g^{(d)}_T(z,\eta, \abs\zeta\abs)+\tilde n_V\, g^{(d)}_V(z,\eta,\abs \zeta\abs),
\end{array}
\ee
with the functions defined below.
For the type II case, one simply takes $\tilde n_T=\tilde n_V=0$.
Note that $p$ is an even function of $\zeta$, as follows from T-duality $R_d \to 1/R_d$.
In this expression, $n_T$ is the number of massless boson/fermion pairs of states in the originally supersymmetric background, for generic $R_d$. $\tilde n_T$ is the number of additional ones at the enhanced gauge symmetry point. The value of $n_V$ is given by the sum over the $n_T$ pairs, with each pair weighted by a sign. The distribution of signs depends on the specific supersymmetry breaking configuration and can yield a negative $n_V$. For the heterotic models we consider, one has
\be
n_T= {2^4\over 2}\, D_0\; ,\qquad  -1\leq {n_V\over n_T}\leq 1\; , \qquad {\tilde n_T\over n_T}={2\over D_0}\; ,\qquad \tilde n_V=\tilde n_T\, .
\ee
The definitions of the various functions appearing in Eq. (\ref{p}) are given by
\be
\!\!\!\!\!\!\!\!\!\!\!\!\!\!\!\!\!\!\!\!\!\!\!\!\!\!\!\!\!\!\!\!\!\!\!\!\!\!\!\!\!\!\!\!\!\!\!\!\!\!\!\!\!\!\!\!\!\!\!\!\!\!\!\!\!\!\!\!\!\!\!\!\!\!\!\!\!\!\!\!\!\!\!\!\!\!\!\!\!\!\!\!\!\!\!\!\!\!\!\!\!\!\!\!\!\!\!\hat f^{(d)}_T(z)=\displaystyle{\Gamma\left({d+1\over 2}\right)\over \pi^{d+1\over 2}}\sum_{\tk_0,\tk_9}{e^{dz}\over \left[e^{2z}(2\tk_0+1)^2+(2\tk_9)^2\right]^{d+1\over 2}}\, ,
\nonumber
\ee
\be
k^{(d)}_T(z,\eta-\abs\zeta\abs)=\displaystyle {\sum_{m_d}}'\abs m_d\abs^{d+1\over 2}e^{{d+1\over 2}(\eta-\abs\zeta\abs)}e^{dz} \sum_{\tk_0,\tk_9}{2K_{d+1\over 2}\left(2\pi\abs m_d\abs e^{\eta-\abs\zeta\abs}\sqrt{e^{2z}(2\tk_0+1)^2+(2\tk_9)^2}\right)\over \left[e^{2z}(2\tk_0+1)^2+(2\tk_9)^2\right]^{d+1\over 4}}\, , \nonumber
\ee
\be
g^{(d)}_T(z,\eta,\abs\zeta\abs)=\displaystyle \left(e^{2\abs\zeta\abs}-1\right)^{d+1\over 2} e^{{d+1\over 2}(\eta-\abs\zeta\abs)}e^{dz}\sum_{\tk_0,\tk_9}{2K_{d+1\over 2}\left(2\pi (e^{2\abs\zeta\abs}-1)e^{\eta-\abs\zeta\abs}\sqrt{e^{2z}(2\tk_0+1)^2+(2\tk_9)^2}\right)\over \left[e^{2z}(2\tk_0+1)^2+(2\tk_9)^2\right]^{d+1\over 4}}\, , \nonumber
\ee
\be
\label{fkgjT}
\!\!\!\!\!\!\!\!\!\!\!\!\!\!\!\!\!\!\!\!\!\!\!\!\!\!\!\!\!\!\!\!\!\!f^{(d+1)}_T(z,\eta-\abs\zeta\abs)=\displaystyle{\Gamma\left({d\over 2}+1\right)\over \pi^{{d\over 2}+1}}\sum_{\tk_0,\tk_9,\tm_d}{e^{(d+1)z}\over \left[e^{2z}(2\tk_0+1)^2+(2\tk_9)^2+e^{-2(\eta-\abs\zeta\abs)}\tm_d^2\right]^{{d\over 2}+1}},
\ee
where $K_\alpha(z)$ are modified Bessel functions of the second kind.
The remaining functions with lower index $V$ are related to those with lower index $T$ by $M\leftrightarrow T$ duality transformations ($z\leftrightarrow -z$):
\be
\label{fkgjV}
\begin{array}{ll}
&\hat f^{(d)}_V(z)=e^{(d-1)z}~\hat f^{(d)}_T(-z) \\\\
&k^{(d)}_V(z,\eta-\abs\zeta\abs)=e^{(d-1)z}~k^{(d)}_T(-z,\eta-\abs\zeta\abs+z) \\\\
&g^{(d)}_V(z,\eta,\abs\zeta\abs)=e^{(d-1)z}~g^{(d)}_T(-z,\eta+z,\abs\zeta\abs)\\\\
&f^{(d+1)}_V(z,\eta-\abs\zeta\abs)=e^{dz~}f^{(d+1)}_T(-z,\eta-\abs\zeta\abs+z) \, .
\end{array}
\ee

We will first focus on the dynamics of the modulus $R_d$, and so we consider the behavior of $-P=-T^dp(z,\eta,\zeta)$ at fixed $T$, $z$ and $\eta$. Note that when the functions $g^{(d)}_T$ and $g^{(d)}_V$ can be neglected, the pressure only depends on two quantities, $z$ and $\eta-|\zeta|$.


\subsection{The five heterotic effective field theory phases}
\label{het pot}

Considering the effect on a single spectator modulus $R_d$ in the case of the heterotic string, the thermal effective potential $-P$ admits five distinct phases corresponding to  different effective field theories.  The form of the potential is sketched in Fig. \ref{fig_potential} and the phases are summarized as follows:
\begin{itemize}
\item  {\bf I: Higgs phase}
\be\left\abs R_d-{1\over R_d}\right\abs < {1\over R_0} ~{\rm and/or}~{1\over R_9}.
\ee
 This phase contains the stringy extended symmetry point at the self-dual  point $R_d=1$. The appropriate effective field theory description is  in terms of a $d$-dimensional theory of gravity coupled to an $SU(2)$ gauge field.
 We will see that the modulus $R_d$ can be stabilized at the self-dual point which turns out to be the minimum of the effective thermal potential.
Indeed, considering the expression of $p$ in (\ref{p}), the functions $g_T^{(d)}$ and/or  $g_V^{(d)}$ are of the same order as $\hat f_T^{(d)}$ and $\hat f_V^{(d)}$, while $k_T^{(d)}$ and $k_V^{(d)}$ are exponentially small due to the behavior of the modified Bessel functions $K_{d+1 \over 2}$. In particular, one has at the origin $\zeta=\ln R_d=0$ the following behavior,
\be
\label{pzeta0}
p(z,\eta,\zeta=0)= (n_T+\tilde n_T)\, \hat f_T^{(d)}(z)+(n_V+\tilde n_V)\, \hat f_V^{(d)}(z):= \tilde p(z).
\ee
This is precisely the form obtained in \cite{Cosmo-1}, when the dynamics of $R_d$ was ignored \ie $R_d$ was taken to be stabilized close to the string scale. In Eq. (\ref{pzeta0}), the contribution of $n_T$ and $\tilde n_T$ is of the same form since both contributions come from massless states when $\zeta$ is at the enhanced gauge symmetry point $\zeta =0$. Due to the fact that $\tilde n_T$ and $\tilde n_V$ are positive, the extremum of $-P$ at $\zeta=0$ is always a minimum.

\item {\bf II:  The flat potential phase}
$$
 {1\over R_0}~ {\rm and} ~{1\over R_9}< R_d - {1\over R_d},
 $$
 \be
 \label{plat}
~ {\rm and} ~~~~~~ R_d< R_0~ {\rm and} ~R_9.
 \ee
For this range of the modulus, there exists a description in terms of a $d$-dimensional theory of gravity coupled to a $U(1)$ gauge field.  Note that the range of this region grows as $R_0$ and $R_9$ increase.  Modulo exponentially suppressed terms ${\cal O}\left[~{\rm exp}(- {\mu\over T}),~{\rm exp}(- {\mu\over M})~\right]$, the potential for the modulus is flat.
We will see that for certain IBC, the modulus $R_d$ may be frozen to an arbitrary value on this plateau, due to the gravitational friction of the expanding universe. The exponentially suppressed terms are irrelevant in this phase and cannot modify this behavior.

In this range (\ref{plat}),  the contributions of $k_T^{(d)}$ and $k_V^{(d)}$, as well as $g_T^{(d)}$ and $g_V^{(d)}$ are exponentially small compared to $f_T^{(d)}$ and $f_V^{(d)}$, so that  $-P$ is independent of $\zeta=\ln R_d$.   The pressure reproduces the result in $d$ dimensions for $n_T$ massless boson/fermion pairs in the originally supersymmetric model (as opposed to phase I which has $n_T + \tilde n_T$ massless pairs).  Physically, we are away from the enhanced symmetry point and so the previous $SU(2)$ states are no longer massless.  More concretely, along the plateau we have
\be
\label{pplat}
p(z,\eta,\zeta)\simeq n_T\, \hat f_T^{(d)}(z)+n_V\, \hat f_V^{(d)}(z):=\hat p(z)\, .
\ee
Either sign of this quantity is allowed when $n_V<0$. Indeed, considering the large $\abs z\abs$ behavior, (\ref{pplat}) implies
\be
\label{largez}
\begin{array}{ll}
-P &\!\!\! \underset{z\to-\infty}\sim-T^d\, n_T\, e^{-z}\, S^o_{d+1}\underset{z\to-\infty} \to -\infty\, ,\\
&\!\!\! \underset{z\to+\infty}\sim -T^d\,  n_V\, e^{dz}\, S^o_{d+1}\underset{z\to+\infty}\to \sign(-n_V)\infty\, ,
\end{array}
\ee
where $S^o_{d}$ (and $S^{oe}_d$ for later use) is  a constant,
\be
\label{Sconstants}
S_d^o={\Gamma\left({d\over 2}\right)\over \pi^{d\over 2}}\,  \sum_m{1\over \abs 2m+1\abs^d}\; , \qquad S_d^{oe}={\Gamma\left({d\over 2}\right)\over \pi^{d\over 2}}\,  {\sum_m}'{1\over \abs m\abs^d},
\ee
and we see that $-P$ may take any value.

\item  {\bf III: Higher-dimensional phase}
\be
R_0~ {\rm and/or} ~R_9<R_d .
\ee
 For large values of the spectator modulus $R_d$, the appropriate effective field theory description is the $(d+1)$-dimensional theory of gravity. The modulus $R_d$ becomes the $\hat g_{dd}$ component of the string frame metric, $\hat g_{dd}=(2\pi R_d)^2$.

All contributions of $\abs m_d\abs$ in $k_T^{(d)}$ and/or $k_V^{(d)}$ are substantial, and the behavior of $p$ is better understood in terms of its second expression in (\ref{p}).  In addition, $g_T^{(d)}$ and $g_V^{(d)}$ are exponentially small. In particular, for $R^\epsilon_d\gg R_9$ and $R_0$ (which are both $\gg 1$)\footnote{We introduce $\epsilon=\sign(\zeta)$ which is 1 in phase III and $-1$ in the T-dual phase V.}, one has
\be
\label{pzetainf}
p(z,\eta,\zeta)\simeq e^{\abs\zeta\abs-\eta-z}\left(n_T\, \hat f_T^{(d+1)}(z)+n_V\, \hat f_V^{(d+1)}(z)+e^{d(z+\eta-\abs \zeta\abs)}(n_T+n_V){S^{oe}_d\over 4}\right)\, ,
\ee
where the term $\dis e^{d(z+\eta-\abs \zeta\abs)}\equiv (R_0/ R_d^\epsilon)^d$  is power-like subdominant. In this expression, we neglect terms that are exponentially small in $(R_d^\epsilon/R_0)$ or $(R_d^\epsilon/R_9)$. The appearance of the functions $\hat f_T^{(d+1)}$ and $\hat f_V^{(d+1)}$ confirms that it is more natural to consider the system in $d+1$ dimensions. This is the case since, in this limit, the circle of radius $R_d$ is very large. In Fig. \ref{fig_potential}, the exponential growth of $-P$ when $\zeta\to + \infty$ is decreasing. This is always the case when $n_V>0$ but for $n_V<0$ is only true when $z$ \ie $M/T$ is small enough. This can be seen by considering the large $\abs z\abs$ limit of (\ref{pzetainf}),
\be
\label{largez'}
\begin{array}{ll}
-P &\!\!\! \underset{z\to-\infty}\sim-T^d\, e^{\abs\zeta\abs-\eta}\, n_T\, e^{-2z}\, S^o_{d+2}  \underset{z\to-\infty}\to -\infty\, ,\\
&\!\!\! \underset{z\to+\infty}\sim -T^d\, e^{\abs\zeta\abs-\eta}\,  n_V\, e^{dz}\, S^o_{d+2}\underset{z\to+\infty}\to \sign(-n_V)\infty\, ,
\end{array}
\ee
where $S^o_{d+2}$ is defined in (\ref{Sconstants}). We will see that the attraction to the RDS$^{d+1}$ implies $z$ to evolve such that the potential for $\zeta$ ends by being exponentially decreasing.  The RDS$^{d+1}$ will then correspond to a run away behavior $R_d(t)\to +\infty$. The RDS$^{d+1}$ is stable only when the subdominant term $(R_0/ R_d^\epsilon)^d$ can be neglected.  This is always true when $R_d$ is macroscopic, since we restrict our study to temperatures above the electroweak scale \ie $R_0$ not very large compared to the internal space and $R_9$ in particular.

If $R_d$ is internal and the subdominant term is not negligible, we will find that the universe is attracted back to phase II, where $R_d$ becomes static and the evolution becomes that of an RDS$^d$.  Finally we note that at early times, where $T$ and $M$ are well above the electro-weak scale, it is possible to have $R_d$ internal while keeping $(R_0/ R_d^\epsilon)^d$ negligible for a large amount of time.  In this case at early times the evolution is initially attracted to an RDS$^{d+1}$.  However, at late times $(R_0/ R_d^\epsilon)^d$ will always become relevant and the evolution always ends in an RDS$^d$.

\item {\bf IV: T-dual flat potential phase}
$$
 {1\over R_0}~ {\rm and} ~{1\over R_9}<  {1\over R_d}-R_d ,
 $$
 \be
 \label{platdual}
~ {\rm and} ~~~~~~ {1\over R_d}< R_0~ {\rm and} ~R_9.
 \ee
The effective theory description is T-dual to the phase II  where the light degrees of freedom are the winding modes  instead of the Kaluza-Klein momentum modes of the phase II.

\item {\bf V: T-dual higher-dimensional phase}
\be
R_0~ {\rm and/or} ~R_9<{1\over R_d} .
\ee
This phase is the T-dual of phase III. The light degrees of freedom are the winding modes.
$-P$ has the same form as in case III, after one transforms $\zeta \rightarrow - \zeta$.  The effective field theory is naturally described in $d+1$ space-time dimensions, with $\hat g_{dd}=(2\pi/R_d)^2$ in string frame.

\end{itemize}

We have to stress here that the above phases arise from a common string setting  but {\it cannot be described in the context of a single field theory}, due to the lack of the string winding modes.  In field theory, only the phases II and III (or their T-dual IV and V) can be described by a single field theory.
In phase I, the string winding modes play a particularly important role for the stabilization of the radius $R_d$  at the extended symmetry point, as shown in Sect. \ref{Higgs}.  In contrast to field theory, string theory naturally interpolates between these various phases.


\subsection{The type II field theory phases}
\label{II pot}

The perturbative type II structure of $-P$ can be derived from the heterotic one by taking $\tilde n_T=\tilde n_V=0$, so that phase I is now equivalent to phases II and III. The local minimum of $-P$ at $\zeta=0$ is not  present anymore and there is a single plateau I $\cup$ II $\cup$ IV (see Fig. \ref{fig_potential}). In phase III (or V), when $R_d^\epsilon \gg R_9$ and $R_0$ (which are both $\gg 1$), the function $p$ in type II is identical to the heterotic one given in Eq. (\ref{pzetainf}). Thus, in type II, the field $\zeta$ admits flat potential phases II and IV in $d$-dimensions, and the higher-dimensional phases III and  V in $d+1$ dimensions.  Again the higher-dimensional phases III and V are stable only to the extent that we can neglect the $(R_0/ R_d^\epsilon)^d$ contribution in Eq. (\ref{pzetainf}), which is always valid for macroscopic values of $R_d^\epsilon$. However,  when $R_d^\epsilon$ is internal, the final evolution is always attracted to the RDS$^d$ of phase II or IV. In addition, during the times for which $(R_0/R_d)^d$ may be ignored, the universe admits an earlier evolution well approximated by an RDS$^{d+1}$.  Since there is no enhancement of $U(1) \rightarrow SU(2)$, the ``$SU(2)$ Higgs phase'' does not exist perturbatively for type II theories.  However, by heterotic-type II duality, we expect non-perturbative effects which may enhance the $U(1) \rightarrow SU(2)$ and imply  an ``$SU(2)$ phase'' I. The non-perturbative effects can correspond to the addition of branes whose separation is governed by the spectator modulus. This will be discussed in more details in Sect. \ref{discuss}. Alternatively, branes wrapped on a vanishing cycle whose size is fixed by the spectator modulus provide another dual type II  set up.

\section{Radiation\red{-like} dominated solutions (RDS) of the\\ Universe and stabilization of the spectator moduli}
\label{rad eras}

In section \ref{het pot}, five distinct phases for the thermal effective potential $-P$ were identified for the heterotic string.  Here, we analyze in detail the behavior of the system in the first three phases.  The behavior of phases IV and V is found from that of phases II and III by T-duality $R_d \to 1/R_d$ and we do not consider them explicitly. We show that the radius $R_d$ can be constant, either at the minimum of the potential in phase I or at any value along the flat region II. In phase III, $R_d$ initially increases along with the expansion of the space-time.  When the quantities $(R_0/ R_d)^d$ and $(R_9/ R_d)^d$ can be neglected, this evolution continues and is well described by an RDS$^{d+1}$. For values of $R_d$ which are internal, $R_d$ is always caught by $R_0$ and $R_9$ and after which the evolution is attracted back to phase II.  A distinct RDS$^d$ exists in regions I, II and IV, while in region III and V there exists an RDS$^{d+1}$ for macroscopic values of $R_d^\epsilon$.

Next we show that these cosmological evolutions are stable against small perturbations.  In particular, for phase I, the spectator modulus $R_d$ is stabilized at the self-dual point, while for phases III (and V) it becomes part of the space-time metric.  For phases II (and IV) the spectator modulus is weakly stabilized due to the presence of gravitational friction arising from the expansion of the universe. From these results, one expects in general that when the dynamics of all spectator moduli are taken into account, the radii that are not dynamically decompactified are (weakly) stabilized at scales smaller than the ones characterizing the temperature and supersymmetry breaking.

For the type II string case described in Sect. \ref{II pot}, there is no phase I, due to the absence of the heterotic $U(1) \rightarrow SU(2)$ enhancement at the self dual point. The remaining type II phases are identical to the heterotic ones.


\subsection{Case I: Higgs phase }
\label{Higgs}

In this case, the radius $R_d$ is naturally interpreted as a scalar Higgs field for an $SU(2)$ gauge group coupled to gravity in $d$ dimensions. As mentioned before, our analysis is restricted to field configurations which are isotropic and homogeneous. We thus look for extrema of the action (\ref{Scompact}) whose metric, temperature and scalars satisfy the ansatz
\be
\label{ansatz d}
ds^2= -dt^2+a(t)^2\left( \left(dx^1\right)^2+\cdots+\left(dx^{d-1}\right)^2\phantom{\dot\Phi}\!\!\!\!\right)\, , \quad T(t)\, , \quad z(t)\, , \quad  \phi_\bot (t)\, ,\quad  \zeta(t)\, .
\ee
The Einstein equations involve a thermal energy-momentum tensor whose components are the energy density $\rho$ and pressure $P$. Using (\ref{pand}) and (\ref{TEMT2}), the energy density $\rho$ takes a factorized form
\be
\label{andrho}
\rho=T^d\, r(z,\eta,\zeta) \qquad \with \qquad r = (d-1) \, p - p_z,
\ee
where  $p_z$ denotes the partial derivative with  respect to $z$.
Given solutions to the scalar equations of motion, we may always find corresponding solutions to the Einstein equations.  We therefore focus first on solving the scalar equations.  Their reduction on the ansatz (\ref{ansatz d}) is given in Eqs. (\ref{eq3o}), (\ref{eq4o}) and (\ref{eq5o}) and summarized here as
\ba
\label{eq3oo}
&&{\cal G}(z,\phi_\bot,\zeta;\o z, \o \phi_\bot, \o \zeta ; \oo z, \oo \phi_\bot, \oo \zeta) + V_z =0\\
\label{eq4oo}
&&h\,\oo\phi_\bot+{1\over d-2}\, (r-p)\, \o\phi_\bot-{1\over \sqrt{d-1}}\, p_\eta=0\\
\label{eq5oo}
&&h\,\oo\zeta+{1\over d-2}\, (r-p)\, \o\zeta- p_\zeta=0,
\ea
where $h$ is defined in Eq. (\ref{h}) and ${\cal G}$ is a function which vanishes when all first and second derivatives in its arguments vanish.  We have reparameterized our fields in terms of the scale factor $\ln a$ so that time-derivatives have been replaced with $(\ln a)$-derivatives denoted as $\o f$.


\subsubsection{Radiation\red{-like} dominated solution}

To start off, we note that the fact the model is invariant under the T-duality $R_d\to 1/R_d$ implies $p(z,\eta,\zeta)$ is an even function of $\zeta$, so that the first derivative of $p$ with respect to $\zeta$ vanishes at $\zeta=0$. Thus, $\zeta \equiv 0$ is a solution to Eq. (\ref{eq5oo}).  Next, from Eq.  (\ref{pzeta0}), the source $P$ is independent of $\eta$ at $\zeta = 0$, so that $p_\eta (z,\eta,0)\equiv 0$. As a consequence, Eq. (\ref{eq4oo}) is solved for any constant $\phi_\bot\equiv \phi_{\bot 0}$, and we find that $\phi_\bot$ remains a modulus.

It is convenient to introduce the quantities $\tilde p(z)$ and $\tilde r(z)$, which are related to the pressure $P$ and energy density $\rho$ at $\zeta=0$ as
\bea
P(z,\eta,0) = T^d p(z,\eta,0) = T^d \tilde p(z) \qquad \qquad
\rho(z,\eta,0) = T^d r(z,\eta,0) = T^d \tilde r(z).
\eea
Eq. (\ref{eq3oo}) implies the complex structure $z$ can be a constant, $z\equiv \tilde z_c$, as long as $\tilde z_c$ is a root of $V_z$. As in Eq. (\ref{simpvz}), here $V_z$ takes the simple form
\be
\label{hatV_z}
\tilde V_z(z):=  V_z(z,\eta,0)= \sqrt{d-2\over d-1}\, (\tilde r-d \, \tilde p)\, .
\ee
The shape of $\tilde V(z)$ depends drastically on the model dependent parameter  $\dis {n_V+\tilde n_V\over n_T+\tilde n_T}\in [-1,1]$ and can be inferred from the behavior for large positive or large negative $z$,
\be
\label{Vzlarge}
\begin{array}{ll}
\tilde V(z)&\!\!\! \dis \underset{z\to-\infty}\sim -e^{(d-1)z}\left( {1\over 2^d-1}+{n_V+\tilde n_V\over n_T+\tilde n_T}\right) \times (n_T+\tilde n_T)\, {d\over 2(d-1)}\sqrt{d-2\over d-1}\, S_d^o\, ,\\\\
&\!\!\! \dis \underset{z\to+\infty}\sim -e^{d \, z}(n_V+\tilde n_V)\times \left(1+{1\over d}\right)\sqrt{d-2\over d-1}\, S_{d+1}^o,
\end{array}
\ee
with $S_{d}^o$ defined  in Eq. (\ref{Sconstants}).  Three cases arise:
\begin{itemize}
\item {\em Case ($\tilde a$)}: For $\dis {n_V+\tilde n_V\over n_T+\tilde n_T}<-{1\over 2^d-1}$, $\tilde V$ increases  monotonically.
\item {\em  Case ($\tilde b$)}: For $\dis -{1\over 2^d-1}<\dis {n_V+\tilde n_V\over n_T+\tilde n_T}<0$, $\tilde V$ has a unique minimum $\tilde z_c$, and $\tilde p(\tilde z_c)>0$.
\item {\em  Case ($\tilde c$)}: For $0<\dis {n_V+\tilde n_V\over n_T+\tilde n_T}$, $\tilde V$ decreases  monotonically.
\end{itemize}
We choose to concentrate on models where $z$ can be stabilized. This corresponds to the {\em  Case ($\tilde b$)} \cite{Francois}\footnote{The models in  {\em  Case ($\tilde c$)} admit a so-called ``Moduli Dominated Solution" corresponding to a contracting Universe (where $z(t)\to +\infty$ is running away) \cite{Francois}. The models in  {\em  Case ($\tilde a$)} are analyzed in \cite{N=1Estes} and admit an RDS in $d+1$ dimensions, after dynamical decompactification of the internal radius $R_9$ involved in the supersymmetry breaking (\ie $z\ll -1$).}, so that
\be
\label{cas b}
\mbox{{\em  Case ($\tilde b$)}}\; : \quad -{1\over 2^d-1}<{n_V+\tilde n_V\over n_T+\tilde n_T}<0,
\ee
which guarantees the possibility to fix $z$ at the critical value $\tilde z_c$ such that (see Eq. (\ref{hatV_z})),
\be
\label{state eq}
\tilde r(\tilde z_c)=d\, \tilde p(\tilde z_c).
\ee
This is the state equation for radiation in $d+1$ dimensions\footnote{As explained in \cite{Cosmo-1, Cosmo-2}, once taking into account the kinetic energy density of the scalar $\Phi$, one recovers the state equation for radiation in $d$ dimensions, as expected for an RDS$^d$.}.  The scalar equations of motion have now been satisfied and the remaining Einstein equations are easily solved.  The overall dependence on time is determined from the Friedmann equation (\ref{h}) which takes the form
\be
\label{rad}
{1\over 2}(d-2)(d-1)\, H^2={\tilde c_r\over a^d}\qquad \where \qquad \tilde c_r={(d-1)^2\over d(d-2)}\, \tilde r(\tilde z_c)\, e^{-d \, \tilde z_c}\, (\tilde a_0 \tilde M_0)^d\, ,
\ee
and is easily integrated. Here, $\tilde a_0$ and $\tilde M_0$ are integration constants.  Using the remaining Eq. (\ref{aTcst}), the full solution we find is,
\be
\label{era}
\begin{array}{c}
\displaystyle a(t)=\left({t\over \tilde t_0}\right)^{2/ d}\, \tilde a_0 \qquad \where \qquad \displaystyle {\tilde a_0\over \tilde t_0^{2/ d}}=\left({d^2\, \tilde c_r\over 2(d-2)(d-1)}\right)^{1/ d}\, ,\\\\
\displaystyle T(t)=M(t)\, e^{-\tilde z_c} = {1\over a(t)}\, e^{-\tilde z_c}\tilde a_0\tilde M_0\; , \qquad \phi_\bot (t)=\phi_{\bot 0}\; ,\qquad \zeta(t)=0,
\end{array}
\ee
 where $\Phi$ is related to $M$ by the definition (\ref{TMz}).
This particular evolution is characterized by a temperature $T$, a spontaneous supersymmetry breaking scale $M$, and an inverse scale factor that are proportional for all times. From the form of the Friedmann equation, Eq. (\ref{rad}), this solution is to be interpreted as a radiation\red{-like} era in $d$ dimensions, with frozen internal radius $R_d\equiv 1$ and a modulus $\phi_\bot$.  In the next sub-section, our aim is to analyze the stability of this heterotic solution and show that it is an attractor of the dynamics for an open set of generic initial boundary conditions.


\subsubsection{Attraction to the radiation\red{-like} era and spectator modulus stabilization}
\label{att d}

To analyze the stability of the radiation\red{-like} era (\ref{era}), we consider small fluctuations around it,
\be
z=\tilde z_c+\ve_{(z)}\; , \qquad  \phi_\bot=\phi_{\bot0}+\ve_{(\phi_\bot)} \; , \qquad\zeta=0+\ve_{(\zeta)}\, ,
\ee
where $\abs\ve_{(z)}\abs$, $\abs\ve_{(\phi_\bot)}\abs$ and $\abs\ve_{(\zeta)}\abs$ are $\ll 1$. The equations of motion for the scalars given in Eqs. (\ref{eq3o}), (\ref{eq4o}) and (\ref{eq5o}) become at first order,
\ba
\label{small z}
&&\!\!\!\!\!\!\!\!\!\!\!\!\!\!\!\!\!\!\!\oo\ve_{(z)}+{1\over 2}\, (d-2) \, \o\ve_{(z)}+\tilde C\, \ve_{(z)}=0\quad \where \quad \tilde C={1\over 2}\, (d-2)^2(d+1)\, \left.{\tilde r_z-d\, \tilde p_z\over d\, \tilde r-\tilde r_z}\right\abs_{\tilde z_c},\\
\label{small ortho}
&&\!\!\!\!\!\!\!\!\!\!\!\!\!\!\!\!\!\!\!\oo\ve_{(\phi_\bot)}+{1\over 2}\, (d-2) \, \o\ve_{(\phi_\bot)}=0\, ,\\
\label{small zeta}
&&\!\!\!\!\!\!\!\!\!\!\!\!\!\!\!\!\!\!\!\oo\ve_{(\zeta)}+{1\over 2}\, (d-2) \, \o\ve_{(\zeta)}+\tilde E(\eta)\, \ve_{(\zeta)}=0\quad \where\quad \tilde E(\eta)=-{(d-2)^2\over 2(d-1)}\, {p_{\zeta\zeta}(\tilde z_c,\eta,0)\over \tilde p(\tilde z_c)}\, .
\ea
It is important to note that even if $p_\zeta$ vanishes and $p$ is independent of $\eta$ when $\zeta=0$,  this is not the case for $p_{\zeta\zeta}$, and indeed we find
\be
\label{pzetazeta}
p_{\zeta\zeta}(z,\eta,0)=32\pi^2\, e^{2(\eta+z)}\left(\tilde n_T\, \hat f_T^{(d-2)}(z)+\tilde n_V\, \hat f_V^{(d-2)}(z)\right)\, .
\ee

Due to the friction term $\half (d-2)$, the solutions of Eq. (\ref{small z}) for arbitrary IBC converge to 0 as $t\to+\infty$ (with eventually damped oscillations), if and only if $\tilde C>0$. This is precisely the case when the condition (\ref{cas b}) is satisfied.  Similarly, all solutions to Eq. (\ref{small ortho}) also converge to 0. Finally, the generic solution of Eq. (\ref{small zeta}) is
\be
\ve_{(\zeta)}=\left( {a_1\over a}\right)^{d-2\over 4}\left[ c_+\, J_{d-1\over 4}\left({d-1\over d-2}\, \left( {a\over a_1}\right)^{d-2\over d-1}\right)+ c_-\, J_{-{d-1\over 4}}\left({d-1\over d-2}\, \left( {a\over a_1}\right)^{d-2\over d-1}\right)\right]\, ,
\ee
where $J_\alpha$ are Bessel functions of the first kind, while $c_+$, $c_-$ (and $a_1$) are constants determined by the IBC. As $t\to +\infty$ or equivalently $a\to +\infty$, the above solution converges to 0 with damped oscillations, due to the behavior of the Bessel functions. We thus conclude that the spectator modulus is stabilized at the self-dual point and that the radiation\red{-like} era (\ref{era}) is a local attractor.


\subsection{Case II: Flat potential phase}
\label{plateau}

We now analyze the phase II.  In this case, $\zeta$ is a flat direction of the thermal effective potential $-P$ which takes the simple form given in Eq. (\ref{pplat}).  There is no enhancement of the gauge group to $SU(2)$ and the spectator modulus $R_d$ is simply a flat direction.  Although there is no potential for $\zeta$ in this phase, fluctuations in $\zeta$ are still suppressed due to the gravitational friction caused by the expansion of the universe.  In addition, we are going to see that depending on the IBC, even if the evolution starts in phase II, it may exit from it and enter either phase I or phase III.  This result is non-trivial due to the fact that as the universe expands, so does the size of the plateau of the flat potential phase.

As in Sect. \ref{Higgs}, we restrict our analysis to isotropic and homogeneous $d$-dimensional universes. We take the same ansatz as in (\ref{ansatz d}), and consequently all of the equations of motion of appendix \ref{eq d} are valid in the case considered here. On the plateau, $p(z,\eta,\zeta)$ is equal to $\hat p(z)$ given in Eq. (\ref{pplat}), where the dependencies on $\eta$ and $\zeta$ are exponentially small in $R_0$ or $R_9$ and are thus neglected. As a consequence, $p_\zeta(z,\eta,\zeta_0)\simeq p_\eta(z,\eta,\zeta_0)\simeq 0$, for any $\zeta_0$ on the plateau.


\subsubsection{Radiation\red{-like} dominated solution}
\label{constant}

Up to the replacement $\zeta(t)\equiv 0 \to \zeta(t)\equiv \zeta_0$, the vanishing of $p_\zeta(z,\eta,\zeta_0)$ and $p_\eta(z,\eta,\zeta_0)$ were the only ingredients used at the beginning of Sect. \ref{Higgs} to derive the radiation\red{-like} dominated solution in $d$ dimensions (\ref{era}).  Thus, in the present case a similar solution  exists. It is obtained by defining $\hat p$ and $\hat r$ as $\tilde p$ and $\tilde r$, but with $\tilde n_T$ and $\tilde n_V$ set to zero
\bea
\hat p = \tilde p |_{\tilde n_T = \tilde n_V = 0}
\qquad \qquad
\hat r = \tilde r |_{\tilde n_T = \tilde n_V = 0} \, .
\eea
The condition similar to (\ref{cas b}) for the existence of a solution with stabilized $z$ is now,
\be
\label{cas b hat}
\mbox{{\em  Case ($\hat b$)}}\; :\quad -{1\over 2^d-1}<{n_V\over n_T}<0,
\ee
in which case there is a unique $\hat z_c$ satisfying
\be
\label{state eq hat}
\hat r(\hat z_c)=d\, \hat p(\hat z_c)\, .
\ee
In phase II, the radiation\red{-like} era can be written as
\be
\label{era plat}
\begin{array}{c}
\displaystyle a(t)=\left({t\over \hat t_0}\right)^{2/ d}\, \hat a_0 \qquad \where \qquad \displaystyle {\hat a_0\over \hat t_0^{2/ d}}=\left({d^2\, \hat c_r\over 2(d-2)(d-1)}\right)^{1/ d}\, ,\\\\
\displaystyle T(t)=M(t)\, e^{-\hat z_c} = {1\over a(t)}\, e^{-\hat z_c}\hat a_0\hat M_0\; , \qquad \phi_\bot (t)=\phi_{\bot 0}\; ,\qquad \zeta(t)=\zeta_0\, ,
\end{array}
\ee
where $\hat a_0$ and $\hat M_0$ are integration constants, $\hat c_r$ is defined as in Eq. (\ref{rad}) but with ``hat'' quantities and $\Phi$ is related to $M$ by the definition (\ref{TMz}).

\subsubsection{Attraction to the radiation\red{-like} era with $R_d$ constant}

The study of the stability given in  Sect. \ref{att d} can also be applied in the present case. More explicitly, the equations of motion for the perturbations of the scalars become
\ba
\label{small z hat}
&&\!\!\!\!\!\!\!\!\!\!\!\!\!\!\!\!\!\!\!\oo\ve_{(z)}+{1\over 2}\, (d-2) \, \o\ve_{(z)}+\hat C\, \ve_{(z)}=0\quad \where \quad \hat C={1\over 2}\, (d-2)^2(d+1)\, \left.{\hat r_z-d\, \hat p_z\over d\, \hat r-\hat r_z}\right\abs_{\hat z_c},\\
\label{small ortho hat}
&&\!\!\!\!\!\!\!\!\!\!\!\!\!\!\!\!\!\!\!\oo\ve_{(\phi_\bot)}+{1\over 2}\, (d-2) \, \o\ve_{(\phi_\bot)}=0\, ,\\
\label{small zeta hat}
&&\!\!\!\!\!\!\!\!\!\!\!\!\!\!\!\!\!\!\!\oo\ve_{(\zeta)}+{1\over 2}\, (d-2) \, \o\ve_{(\zeta)}=0\, ,
\ea
where the analogue of $\tilde E(\eta)$ in Eq. (\ref{small zeta}) vanishes, since $p_{\zeta\zeta}(\hat z_c,\eta,\zeta_0)\simeq 0$.  The positivity of $\hat C$ is again guaranteed by the condition (\ref{cas b hat}) and all perturbations $\ve_{(z)}$, $\ve_{(\phi_\bot)}$, $\ve_{(\zeta)}$ converge to 0 as $t\to +\infty$.

We conclude that if $\zeta$ is initially in the range where its potential is flat, some generic IBC imply an attraction to a regime where $\zeta$ and $\phi_\bot$ are constant moduli. Even if the potential of the spectator modulus $R_d$ and $\phi_\bot$ are flat, these fields are nevertheless ``stabilized in a weak sense'' since the expansion of the universe dilutes their kinetic energies which then vanish for late times.
Physically, the friction terms proportional to $H$ freeze them in place at late times.
After this has occurred, $R_d(t)$ is a constant, while $R_9(t)$ and $R_0(t)$ behave as  $a^{d-2\over d-1}$ and are thus increasing such that the complex structure-like ratio $e^z=R_0(t)/R_9(t)$ is stabilized.


\subsubsection{Falling off the plateau}
\label{fall}

We now show that despite the expansion of the plateau and the friction terms, for sufficiently large initial velocities, it is always possible for $\zeta$ to escape from the plateau and enter regions III or I.  To show this, we will simply find particular IBC that yield this precise behavior.

Suppose $\zeta$ is somewhere along the plateau and choose, as in Sect. \ref{constant}, $z\equiv \hat z_c$ and $\phi\equiv \phi_{\bot 0}$, which  are trivial  solutions to Eqs (\ref{eq3oo}) and (\ref{eq4oo}). Eq. (\ref{oPhi}) gives $\o \Phi=-\sqrt{d-2\over d-1}$, so that the definition (\ref{Phiphibot}) implies
\be
\o \eta=(\o \eta+\o z) ={d-2\over d-1}\, ,
\ee
which is nothing but the ``constant velocity'' of the right edge of the plateau. This defines an escape velocity and we now wish to know if it is possible for $\o \zeta$ to be larger than this speed for a long enough time to reach the edge of the plateau.
One can see that taking $\zeta$ as,
\be
\label{zeta roll}
\zeta=\zeta_0+(d-2)\sqrt{d\over d-1}\, \ln a \, ,
\ee
solves the $\zeta$-equation of motion (\ref{eq5oo}).\footnote{One may worry that this solution yields a singular $H$.  To remedy this, one can introduce a perturbation $\o\ve_{(\zeta)}:=c\, a^{d-2}$ to $\zeta$,  such that $|c a_{\mbox{\tiny fall}}^{d-2}| \ll 1$, where $a_{\mbox{\tiny fall}}$ is defined in Eq. (\ref{afall}).}
 We see that $\zeta$ is rolling at approximately (see the previous footnote) ``constant velocity'' given by $\o \zeta \simeq (d-2)\sqrt{d\over d-1}$, and in particular we have $\o \zeta > \o \eta$.  As a result, for initial values $\zeta_{\mbox{\tiny init}}$,  $\eta_{\mbox{\tiny init}}$ and $a_{\mbox{\tiny init}}$, there is always a scale factor $a_{\mbox{\tiny fall}}$ where $\zeta$ reaches the right boundary of the plateau. It is defined by
\be
\label{afall}
\min(\eta_{\mbox{\tiny init}}, \eta_{\mbox{\tiny init}}+z_c)-\zeta_{\mbox{\tiny init}}=\left((d-2)\sqrt{d\over d-1}-{d-2\over d-1}\right) \left( \ln a_{\mbox{\tiny fall}} - \ln a_{\mbox{\tiny init}}\right)>0\, .
\ee

Another solution with opposite velocity $\o\zeta$ is also allowed, so that $\zeta$ can roll and enter  phase I.


\subsection{Case III: Higher-dimensional phase}
\label{spatial}

Here we analyze phase III of Sect. \ref{het pot}.  When $R_d$ is comparable to the size of $T^{d-1}$, it is natural to interpret the model in $d+1$ dimensions. The fields with non-trivial vacuum expectation value are the Einstein frame metric $g'_{\mu\nu}$ and dilaton $\dil'$ in $d+1$ dimensions, coupled to the scalar $\eta$ defined in (\ref{scalars}). We use primes here to denote that the fields are now normalized in $d+1$ dimensions and are not the same as the fields appearing in the two previous sections.  Note that  $\zeta =\ln R_d$ is now one of the degrees of freedom of $g'_{\mu\nu}$. In the notations of appendix \ref{eq d+1}, we are now in the case $n=1$ and $\Delta = 8-A-d$ (where $A=0$ for toroidal models and $A=4$ for orbifold ones), so that the effective space-time dimension is ${\cal D} = d+1$.

The action $S$ in (\ref{Scompact}) can be written equivalently as
\be
\label{Sdecompact}
S=\int d^{d+1}x\sqrt{-g'}\left( {R'\over 2}-{1\over 2} (\partial \Phi')^2-{1\over 2}(\partial \phi'_\bot)^2+P'\right),
\ee
where we have introduced the canonically normalized fields
\be
\Phi':={2\over \sqrt{d(d-1)}}\, \dil'-\sqrt{d-1\over d}\, \eta\; , \qquad \phi'_\bot:={2\over \sqrt{d}}\, \dil' +{1\over \sqrt{d}}\, \eta\, .
\ee
The pressure $P'$ defined in $d+1$ dimensions is the free energy density $-{\cal F'}$ (see Eq. (\ref{p=-F})). It is related to the partition function $Z$ by,
\be
P'=e^{{2(d+1)\over d-1}\dil'}\, {Z\over V_{0,...,d}}\, ,
\ee
where the partition function is given as before in Eq. (\ref{Z}).
In $d+1$ dimensions, the temperature $T'$ and supersymmetry breaking scale $M'$ (both measured in Einstein frame, see Eqs (\ref{TD}) and (\ref{MD}))  have different normalizations compared to their counterparts in Eqs (\ref{TMz}), but the complex structure $z$ remains the same,
\be
\label{T'M'z}
T'={e^{{2\dil'\over d-1}}\over 2\pi R_0\, }\; , \qquad M'={e^{{2\dil'\over d-1}}\over 2\pi R_9\, }\equiv{e^{\sqrt{d\over d-1}\Phi'}\over 2\pi}\; , \qquad e^z:={M'\over T'}={R_0\over R_9}\, .
\ee

We are interested in extrema of $S$ that can be interpreted as homogeneous but anisotropic space-times with rotation group $SO(d-1)$. We consider the ansatz
\be
\label{ansatz d+1}
ds^{\prime 2}= -dt^2+a'(t)^2\left( \left(dx^1\right)^2+\cdots+\left(dx^{d-1}\right)^2\phantom{\dot\Phi}\!\!\!\!\right)+ b(t)^2\left(dx^d\right)^2\, , \quad T'(t)\, , \quad \Phi'(t)\, , \quad  \phi'_\bot (t)\, ,
\ee
where the metric scale factor  $b$ along the direction $d$ is related to $\zeta$ by\footnote{We introduce absolute values $\abs\zeta\abs\equiv\epsilon\zeta$ for our analysis to also be valid in the T-dual phase V.}
\be
\label{b}
b:=e^{\abs\zeta\abs-{2\dil'\over d-1}} \, .
\ee
The thermal part of the energy-momentum tensor involves an energy  density  $\rho'$, the pressure $P'$ (associated to the directions $1,\dots,d-1$) and a pressure $P'+b(\partial P'/\partial b)$ in the direction $d$ (see Eqs (\ref{TEMT2})--(\ref{P4})). They are more conveniently written as functions of $\{T',z,\eta,\zeta\}$ in terms of which they take a factorized form,
\be
P'\equiv T^{\prime d+1}\, p'(z,\eta,\abs\zeta\abs)\; ,
\qquad \qquad
\rho'\equiv T^{\prime d+1}\, r'(z,\eta,\abs\zeta\abs)\, .
\ee
The functions $p'$ and $r'$ are related to their counterparts in $d$ dimensions $p$ and $r$ (Eqs. (\ref{pand}) and (\ref{andrho})) as,
\be
\label{p'}
p'(z,\eta,\abs\zeta\abs)= e^{\eta-\abs\zeta\abs+z}\, p(z,\eta,\abs\zeta\abs),
\quad r'(z,\eta,\abs\zeta\abs)= e^{\eta-\abs\zeta\abs+z}\, r(z,\eta,\abs\zeta\abs), \quad
r'=d \, p'- p_z'.
\ee

The equations of motion reduced on the ansatz (\ref{ansatz d+1}) are derived in appendix \ref{eq d+1}. There are three independent Einstein equations, coupled to two scalar equations. It is relevant to introduce a modulus $\xi$ as
\bea
e^\xi := {b \over  a'}\, ,
\eea
which is a ``complex structure for the external space''. Then, one can replace one of the Einstein equations by the equation for the scalar $\xi$. Effectively, we have three scalars, whose equations of motion (\ref{eq'4o}), (\ref{eq'5o}) and (\ref{eq'3''o}) can be solved, before considering the remaining Einstein equations. The scalar equations of motion are summarized here as
\bea
\label{eq'4oo}
&&{\cal G}'(z,\phi_\bot', \zeta;  \o z, \o \phi_\bot{\!\!\!}'\,\,, \o \zeta;\oo z, \oo \phi_\bot{\!\!\!}'\,\,, \oo \zeta) + V_z' =0\, ,\\
\label{eq'5oo}
&&h'\, \oo\phi_\bot{\!\!\!}'\,\,+{1\over d-1}\, \left(r'-p'-p'_{\abs\zeta\abs}\right)\, \o\phi_\bot{\!\!\!}'\,\,-{1\over \sqrt{d}}\left(p'_\eta+p'_{\abs\zeta\abs}\right)=0\, ,\\
\label{eq'3''oo}
&&h'\, \oo\xi+{1\over d-1}\, \left(r'-p'-p'_{\abs\zeta\abs}\right)\, \o\xi- p'_{\abs\zeta\abs}=0\, ,
\eea
where $h'$ is defined in Eq. (\ref{h'}) and ${\cal G}'$ is a function which vanishes when all its first and second derivatives in its arguments vanish.


\subsubsection{Radiation\red{-like} dominated solution in $d+1$ dimensions}
\label{sol d+1}

Our aim is to study the dynamics when the characteristic size of the direction $d$ is larger than the scale of the internal space. When $R_d\gg R_9$ and $R_0$,
one observes from Eqs. (\ref{p'}) and (\ref{pzetainf}) that if we neglect the subdominant term $e^{d(z+\eta-\abs \zeta\abs)}=(R_0/R_d^\epsilon)^d$, the pressure $P'$ is independent of $\eta$ and $\zeta$
\be
p'(z,\eta,\abs\zeta\abs)\simeq n_T\, \hat f_T^{(d+1)}(z)+n_V\, \hat f_V^{(d+1)}(z):=\hat p'(z)\quad \when\quad \abs\zeta\abs\gg\eta~\and~ \eta+z\, .
\ee
 In this regime, we define similarly $\hat r'(z):=r'$ and observe that since $p'_\eta\simeq p'_{\abs\zeta\abs}\simeq 0$, constant $\phi_\bot{\!\!\!}'\,\, \equiv \phi_{\bot 0}{\!\!\!\!}'\,\,$ and $\xi\equiv \xi_0$ solve trivially Eqs (\ref{eq'5oo}) and (\ref{eq'3''oo}).  $\phi_\bot{\!\!\!}'$ and $\xi$ are thus moduli in this limit. Note that neglecting the power-like term $(R_0/R_d^\epsilon)^d$ is justified if $R_d$ is macroscopic, since we restrict our analysis to temperatures above the electroweak scale.  In particular, if $R_d$ takes values such that $(R_d/R_0)^d \gtrsim e^{2\pi R_0}$ and $(R_d/R_0)^d \gtrsim e^{2\pi R_9}$ it is of the order of terms which we have already dropped.

A constant $z\equiv \hat z_c'$ is allowed by Eq. (\ref{eq'4oo}) if $V_z'=0$. In the regime we focus on with $R_d\gg R_9, R_0$, the definition (\ref{V'_z}) simplifies to
\be
V_z'(z,\eta,\abs\zeta\abs)\simeq \sqrt{d-1\over d} \left(\hat r'-(d+1)\, \hat p'\phantom{\dot\Phi}\!\!\!\!\right):= \hat V_z'(z) ,
\ee
which is identical to (\ref{hatV_z}) in $d+1$ dimensions. $\hat V'(z)$ admits a critical point in what we call {\em  Case ($\hat b'$)} (by analogy with (\ref{cas b hat})), defined by the condition
\be
\label{cas b'}
\mbox{{\em  Case ($\hat b'$)}}\; :\quad
-{1\over 2^{d+1}-1}<{n_V\over n_T}<0\, .
\ee
When this is satisfied, $\hat z_c'$ is the unique solution to
\be
\label{state eq'}
\hat r'(\hat z'_c)=(d+1)\, \hat p'(\hat z'_c)\, ,
\ee
which is the state equation of radiation in $d+2$ dimensions\footnote{One has to take into account the classical kinetic energy part of the stress tensor to recover the equation of state for radiation in $d+1$ dimensions.}.

The scalar equations of motion Eqs. (\ref{eq'4oo}), (\ref{eq'5oo}) and (\ref{eq'3''oo}) have now been satisfied.  The remaining Einstein equations are easily integrated.  The overall time dependence is determined by the Friedmann equation (\ref{h'}), which becomes
\be
\label{rad'}
{1\over 2}(d-1)d\, H^{\prime 2}={\hat c'_r\over a^{\prime d+1}}\qquad \where \qquad \hat c'_r={d^2\over (d+1)(d-1)}\, \hat r'(\hat z'_c)\, e^{-(d+1) \hat z'_c}\, (\hat a'_0\hat M'_0)^{d+1}\, .
\ee
The full solution we obtain is,
\be
\label{era'}
\begin{array}{c}
\displaystyle a'(t)=\left({t\over \hat t'_0}\right)^{2\over d+1}\, \hat a'_0 \qquad \where \qquad \displaystyle {\hat a'_0\over {{\hat t}{}'_0}^{2\over d+1}}=\left({(d+1)^2\, \hat c'_r\over 2(d-1)d}\right)^{1\over d+1}\, ,\\\\
\displaystyle T'(t)=M'(t)\, e^{-\hat z'_c} = {1\over a'(t)}\, e^{-\hat z'_c}\hat a'_0\hat M'_0= {1\over b(t)}\, e^{\xi_0-\hat z'_c}a'_0M'_0\; , \qquad \phi'_\bot (t)=\phi'_{\bot 0}\, ,
\end{array}
\ee
where $\Phi'$ is related to $M'$ by definition (\ref{T'M'z}), and $\hat a'_0$, $\hat M'_0$, $\xi_0$ and $\phi'_{\bot 0}$ are arbitrary integration constants. For this particular evolution, it is useful to rescale the space coordinate $x^d$ to bring the metric (\ref{ansatz d+1}) in an isotropic form,
\be
x^{\prime d}:=e^{\xi_0}\, x^d\quad \Longrightarrow\quad ds^{\prime 2}= -dt^2+a'(t)^2\left( \left(dx^1\right)^2+\cdots+\left(dx^{d-1}\right)^2+\left(dx^{\prime d}\right)^2\phantom{\dot\Phi}\!\!\!\!\right).
\ee
This shows that there is an enhancement of the local rotation group, $SO(d-1)\to SO(d)$. We learn from the effective Friedmann equation (\ref{rad'}) that this evolution of the universe can be interpreted as a radiation\red{-like} era in $d+1$ dimensions. The temperature $T'$ and supersymmetry breaking scale $M'$ are inversely proportional to the isotropic scale factor $a'$, while $\phi'_\bot$ remains a modulus. Next, we study the stability and attraction properties of this solution.


\subsubsection{Attraction to the radiation\red{-like} era in $d+1$ dimensions via decompactification of $S^1(R_d)$}
\label{att d+1}

To study the stability of the solution (\ref{era'}) (which is valid when the subdominant term $(R_0/R_d^\epsilon)^d$  is negligible in Eq. (\ref{pzetainf})), we analyze the behavior of the small fluctuations around it,
\be
z=\hat z'_c+\ve_{(z)}\; , \qquad  \phi_\bot'= \phi_{\bot 0}'+\ve_{(\phi'_\bot)} \; , \qquad \xi=\xi_0+ \ve_{(\xi)}\, ,
\ee
where $\abs\ve_{(z)}\abs$, $\abs\ve_{(\phi'_\bot)}\abs$ and $\abs\ve_{(\xi)}\abs$ are $\ll 1$.  The scalar equations of motion, Eqs (\ref{eq'4o}),  (\ref{eq'5o}) and (\ref{eq'3''o}), in this regime become
\ba
\label{small z'}
&&\!\!\!\!\!\!\!\!\!\!\!\!\!\!\!\!\!\!\!\oo\ve_{(z)}+{1\over 2}\, (d-1) \, \o\ve_{(z)}+\hat C'\, \ve_{(z)}=0\, \\
\label{small ortho'}
&&\!\!\!\!\!\!\!\!\!\!\!\!\!\!\!\!\!\!\!\oo\ve_{(\phi'_\bot)}+{1\over 2}\, (d-1) \, \o\ve_{(\phi'_\bot)}=0\, ,\\
\label{small xi}
&&\!\!\!\!\!\!\!\!\!\!\!\!\!\!\!\!\!\!\!\oo\ve_{(\xi)}+{1\over 2}\, (d-1) \, \o\ve_{(\xi)}=0\, ,
\ea
where
\be
\hat C'={1\over 2}\, (d-1)^2(d+2)\, \left.{\hat r'_z-(d+1)\, \hat p'_z\over (d+1)\, \hat r'-\hat r'_z}\right\abs_{\hat z'_c}\, .
\ee
We find that $\hat C'$ always satisfies $\hat C'>0$ in
{\em  Case ($\hat b'$)} defined in (\ref{cas b'}). This implies  that for arbitrary IBC, the solutions of Eq. (\ref{small z'}) converge to 0 (with eventually damped oscillations) as $t\to+\infty$.  In addition, all solutions to Eqs (\ref{small ortho'}) and (\ref{small xi}) also converge to 0, when $t\to +\infty$.

Thus, the radiation\red{-like} era in $d+1$ dimensions (\ref{era'}) is stable under small fluctuations, when the subdominant term $(R_0/R^\epsilon_d)^d$ in the Eq. (\ref{pzetainf}) is neglected. There is an open set of IBC, in particular when $R_d$ is initially macroscopic, such that the solutions are attracted by this evolution, which is characterized by an enhanced local rotation group. The \Ka modulus $R_d$ is better understood in terms of the ``external space complex structure'' ratio $e^\xi=b/a=R_d/R$, where $R$ is the radius of one of the $d-1$ space-like dimensions.  This is similar to what is happening to the \Ka modulus $R_9$ that we consider through the complex structure-like ratio $e^z=M/T=R_0/R_9$. However, while $z$ is dynamically stabilized to the value $\hat z'_c$, $\xi$ is a modulus $\xi_0$. This last remark is due to the fact that we consider local equations of motion only. An additional choice of global boundary conditions on the relative sizes of the large external dimensions would specify $\xi_0$. Only astrophysical observations may be sensitive to moduli such as $\xi_0$, but not ``local'' experiments encountered in particle physics.


\subsubsection{Residual force and attraction back to phase II}
\label{back II}

Here, we want to study the effect of the subdominant term in Eq. (\ref{pzetainf}).  We take it small compared to 1 but not negligible and in particular, we want to analyze how it perturbs the results of the previous subsection, namely the attraction to the RDS$^{d+1}$.
We first introduce $y$ as
\be
e^y:= e^{z+\eta-\abs \zeta\abs}=\left( {R_0\over R_d^\epsilon}\right),
\ee
It will be convenient to express the conservation of the energy-momentum tensor (\ref{aTxicst}) in the following form,
\be
\label{CEy}
e^{dy}=e^{-(d-1)\xi}(r'+p')\times \mbox{cst}.
\ee

Let us consider a generic perturbation around  the RDS$^{d+1}$,
\be
z=\hat z'_c+\ve_{(z)}\; , \qquad  \phi_\bot'= \phi_{\bot 0}'+\ve_{(\phi'_\bot)} \; , \qquad \xi=\xi_0+ \ve_{(\xi)}\; ,\qquad y=y_0+\varepsilon_{(y)},
\ee
where $\abs\ve_{(z)}\abs$, $\abs\ve_{(\phi'_\bot)}\abs$, $\abs\ve_{(\xi)}\abs$, $\abs\ve_{(y)}\abs$ and the constant $e^{dy_0}$ are $\ll 1$. At order one, the equations for $\ve_{(z)}$ and $\ve_{(\phi'_\bot)}$ are identical to (\ref{small z'}) and (\ref{small ortho'}), when $e^{dy_0}$ was neglected. However, the equation for $\ve_{(\xi)}$ becomes
\be
\oo\ve_{(\xi)}+{1\over 2}\, (d-1) \left( \o\ve_{(\xi)}+c_1 \, e^{dy_0}\right) =0\; \qquad \where\qquad  c_1=(n_T+n_V){(d-1)S^{oe}_d\over 4\, \hat p'(\hat z'_c)}>0,
\ee
while Eq. (\ref{CEy}) gives
\be
\o\ve_{(y)}=-c_2\, \o\ve_{(\xi)} +c_3\, \o\ve_{(z)}\quad \where\quad c_2=1-{1\over d}\; , \quad c_3={1\over d}\left.{\hat r_z'+\hat p_z'\over \hat r'+\hat p'}\right\abs_{\hat z'_c}.
\ee
This implies $\o\ve_{(z)}=\o\ve_{(\phi'_\bot)}=0$, $\o \ve_{(\xi)}= -c_1\, e^{dy_0}$ and  $\o \ve_{(y)}= c_1\, c_2\, e^{dy_0}>0$. We conclude that $e^y=R_0/R_d^\epsilon$ and  $e^{y-z}=R_9/R_d^\epsilon$ are slowly increasing \ie the distance between $R_d^\epsilon$ and the plateau decreases. The interpretation of the evolution in terms of RDS$^{d+1}$ remains valid until $e^{dy}$ and $e^{d(y-z)}$ cease to be $\ll 1$. As said in the previous subsection, this cannot happen if $R_d$ is initially macroscopic and we are interested in temperature and supersymmetry breaking scales larger than the electroweak scale.\footnote{In non-realistic models (such as the ones considered in this work) where no new physics arises at the electroweak scale, considering macroscopic but finite $R_0$ and $R_9$ catching $R_d$ would correspond to a phase of the universe in $d+2$ dimensions with the temperature effectively zero.}  However, if initially $R_d^\epsilon$ is larger than $R_0$ and $R_9$ but still microscopic, the above analysis suggests that it will be ``caught by the plateau'' and the cosmology is attracted back to phase II \ie the RDS$^d$.


\section{Non-perturbative cosmologies in type II}
\label{discuss}

As discussed in section \ref{II pot}, the thermal effective potential in type II theories does not give rise to a Higgs phase I as in the heterotic string.  The reason for this difference is that at the self-dual point, there is no enhancement  $U(1) \rightarrow SU(2)$ and thus no growth of the number of massless degrees of freedom.  However, we expect by heterotic-type II duality that such a phase should be possible in type II at the non-perturbative level.  A natural candidate setup to produce this effect in type II is to introduce a pair of D-branes, whose separation is related to the spectator modulus $R_d$.  The stabilization of $R_d$ at the self-dual point in the heterotic case suggests in the dual type II picture that the effect of the thermal effective potential is to fix the D-branes on top of each other, thus producing an $U(1) \rightarrow SU(2)$ enhancement.  This attractive force between the D-branes will only be local, in the sense that if we separate the D-branes from each other so that $R_d$ is in the range of  phase II of Sect. \ref{het pot}, the thermal effective potential should allow stable finite distances between the D-branes.  If we further increase the separation, we expect to reach a point where the thermal effective potential induces a repulsive force that pushes the D-branes away from each other and we are entering phase III. However, $R_0(t)$ and $R_9(t)$ are increasing faster than $R_d(t)$.  When they catch it, the force between the D-branes vanishes and the distance between them becomes static and we are back in phase II.  On the contrary, if the distance between the D-branes is macroscopic, the effective potential induces a repulsive force that pushes the D-branes away from each other and we are in the higher-dimensional phase III.

Another set up dual to the heterotic gauge group enhancement can be considered in terms of singularities in the internal space. For instance, a type IIA D2-brane wrapped on a vanishing $\Co\Pb^1$ cycle of radius dual to $R_d$ can give rise to an $SU(2)$ gauge theory and admits a mirror description in type IIB \cite{het/II}. The equivalence between the brane-world and geometrical singularity pictures can be analyzed along the lines of Ref. \cite{KP}.


\section{Conclusion and discussion}

In this work, we have considered string theory models in flat space, where geometrical fluxes induce a spontaneous breaking of supersymmetry and finite temperature.  We have computed the 1-loop free energy density, which is nothing but the effective potential at finite temperature and first order in perturbation theory. It depends on the temperature $T$, the supersymmetry breaking scale $M$ and ``spectator moduli'' that characterize the internal space but {\em are not} involved in the breaking of supersymmetry. Our aim was to analyze the dynamics of these spectator moduli in the presence of both temperature and supersymmetry breaking.

We have analyzed in many details heterotic and type IIB models where the dynamics of only one of the spectator radii, $R_d$, is taken into account. More precisely, we have considered Euclidean backgrounds which are of the form $S^1 (R_0)\times T^{d-1}\times S^1(R_d)\times {\cal M} \times S^1(R_9)$, where $S^1(R_0)$ and $S^1(R_9)$ both contain fluxes.  The flux along the Euclidean time cycle $S^1(R_0)$ introduces temperature $T\propto 1/R_0$ and the flux along $S^1(R_9)$ implies the spontaneous breaking of supersymmetry at a scale $M\propto 1/R_9$. The torus  $T^{d-1}$ is very large, while the internal manifold ${\cal M}$ is either $T^{8-d}$ or $T^{4-d}\times \dis {T^4\over \Z_2}$, with fixed radii close to the string length.

In heterotic models, we found five distinct phases of the thermal effective potential (see Fig. \ref{fig_potential}). In phase I, the potential plays a role in confining the spectator modulus by giving it an effective mass in addition to the gravitational friction effects. $R_d$ plays the role of a Higgs field stabilized at the enhanced gauge symmetry point $R_d=1$, where $U(1)\to SU(2)$.
In phase II (or IV), $R_d$ converges to an arbitrary constant.  This is simply due to the gravitational friction arising from the expansion of the universe. Thus, while the modulus may take any value, its excitations always die off as the universe expands.  In phase III (or V), if $R_d$ (or $1/R_d$) is macroscopic\red{, meaning that we may always neglect the sub-dominant term in the effective potential (\ref{pzetainf})}, then $R_d$ increases proportionally to the expansion of the universe. In this case, the dynamics of the modulus is better understood in terms of a complex structure characterizing the anisotropy of a $(d+1)$-dimensional universe.  As in phases II and IV, the excitations of this complex structure die off due to gravitational friction. If instead $R_d$ is internal, it always enters phase II (or IV).

The analysis of the type IIB case is qualitatively the same, up to an important difference. The heterotic Higgs phase does not exist, since there is no gauge symmetry enhancement at $R_d=1$ in type II superstrings, at least in a perturbative approach.  However, we expect by heterotic-type II duality that such a gauge theory enhancement should occur once taking into account non-perturbative effects in type II superstrings. In particular, the modulus governing the distance between D-branes or the size of some cycle on which a brane is wrapped could play the dual role of the heterotic radius $R_d$.

The heterotic picture of phase I naturally generalizes to the case where all spectator moduli are allowed to vary.  Although we did not consider this case explicitly, one may analyze models with non-diagonal tori.  We expect the existence of local minima of the thermal effective potential at each enhanced symmetry point, as a consequence of the increase in the number of light states.  The lowest such point should be given by the most symmetric point.

For the backgrounds $S^1 (R_0)\times T^{d-1}\times S^1(R_d)\times {\cal M} \times S^1(R_9)$, the thermal effective potential has a universal form, up to model-dependent integer parameters ($n_T,n_V$). $n_T$ is the number of massless boson/fermions pairs in the original supersymmetric model \ie before the fluxes are switched on. $n_V$ depends on the precise prescription chosen to break supersymmetry. In the heterotic phase I, $(n_T,n_V)$ has to be replaced by  $(n_T+\tilde n_T, n_V+\tilde n_V)$ to account for the additional massless states arising at the enhanced symmetry point. Depending on the IBC, we found that the dynamics of the Universe can be attracted to either of the five Radiation\red{-like} Dominated Solutions associated to the five phases, provided the following conditions are fulfilled:
\be
\label{finalrestrictions}
\begin{array}{lll}
\mbox{phase I} & :\quad&\dis  -{1\over 2^d-1} < {n_V+\tilde n_V\over n_T+\tilde n_T}<0 , \\\\
\mbox{phases II}\cup{\rm IV}&:\quad&\dis  -{1\over 2^d-1} < {n_V\over n_T}<0 , \\\\
\mbox{phases III}\cup{\rm V}&:\quad&\dis  -{1\over 2^{d+1}-1} < {n_V\over n_T}<0.
\end{array}
\ee
When some (or all) of these conditions are not satisfied, different histories of the Universe are possible. For instance, suppose a model satisfies the second of the above conditions, but not the third, with $R_d$ macroscopic and thus initially in phase III. Depending on the remaining initial boundary data, we conjecture at least three different late time behaviors to arise, which again correspond to Radiation\red{-like} Dominated Solutions:
\begin{itemize}
\item A dynamical compactification of the spectator radius $R_d$ to enter phases II, I, or IV. The attractor is an RDS$^d$, where $M\propto 1/R_9$.
\item A dynamical decompactification of the radius $R_9$ that participates in the spontaneous breaking of supersymmetry. This mechanism was conjectured in \cite{Francois} and is shown in an explicit example in \cite{N=1Estes}. The attractor solution is an RDS$^{d+2}$, where supersymmetry is spontaneously broken by thermal effects only (no $M$).
\item A (non-perturbative) connection to a cousin model with flux in some of the previously spectator directions. For instance, for $d+1=4$ and flux in two internal directions, say 8 and 9, the constraint $\dis -{1\over 15}<{n_V\over n_T}<0$ is replaced by the less restrictive one $\dis -0.215 <{n_V\over n_T}<0$ (see \cite{Cosmo-2}). The attractor solution is an RDS$^{d+1}$, where $M\propto 1/\sqrt{R_8R_9}$.  The interesting point here is that the solution is stabilized by the spontaneous generation of topological flux.
\end{itemize}

In this work, we restricted our discussion of orbifold models to cases where $\dis {T^4\over \Z_2}$ did not contain flux and had its radii fixed close to the string scale. In the companion paper \cite{N=1Estes}, we fill this gap and extend the analysis to cases where an orbifold action is non-trivial on dynamical circles and whose radii are either participating in the breaking of supersymmetry or are spectators.

\red{We would like to conclude by giving a summary of the results which have been made in studying the intermediate cosmological regime within the framework of string theory, as developed in \cite{Cosmo-1,Cosmo-2,Francois,Cosmo-0} as well as the current paper:
\begin{itemize}
\item The first result is the discovery of the Radiation-like Cosmological Solutions ($RDS^d$). These $RDS^d$ solutions are not the ``usual" radiation solutions defined by $\rho_{\rm thermal}=(d-1)P_{\rm thermal}$, but instead satisfy $\rho_{\rm total}=(d-1)P_{\rm total}$ only after the contribution from the coherent evolution of the supersymmetry breaking modulus is included.  These solutions, are also consistent in that the evolution of the space-time curvature scales, ${H^2=(\dot a}/a)^2$ and $\dot H$, the dilaton and the evolving radii scales, $({\dot \phi}_{\mbox{\tiny dil}})^2$, ${\ddot \phi}_{\mbox{\tiny dil}}$,$({\dot R}_I/R_I)^2$, $\ddot R_I/R_I$, is such that they decrease as time progresses.  Thus if one starts with small curvature and small coupling, at later times one is guaranteed to remain within the regime of small curvature and small coupling.
\item The second result is the discovery of the ``attractor mechanism" which is valid in the intermediate cosmological regime (after the Hagedorn transition or alternatively the inflation era but before electro-weak symmetry breaking); within this era the $RDS^d$ cosmological solutions are not only stable under small fluctuations but also are the only solutions (``attractors") at late cosmological times. Furthermore, thanks to this attractor mechanism, most of the Hagedorn exit ambiguities are washed out in later cosmological times.
\item The third result, is the fact that it is possible to derive at the string perturbative level, (however exact in $\alpha'$), the full string free energy  ${\cal F}(T,M; \mu_I)$ as a functional of all moduli (including the SUSY breaking moduli $T,M$, the string coupling constant modulus $\dil$ as well as  the ``spectator moduli" $\mu_I$),  at least for a certain class of string vacua where the spontaneous breaking of supersymmetry is induced by geometrical fluxes.
\end{itemize}}

\red{There are two main directions in going further with this approach.  First,}
one may carry out our analysis with four-dimensional heterotic models, whose internal space $\dis {T^6\over \Z_2\times \Z_2}$ in the presence of fluxes breaks spontaneously $\N_4=1$  supersymmetry.
Depending on the details of the internal space and spontaneous supersymmetry breaking configuration, it is possible for an additional scale $Q$ to appear at very late cosmological  times. $Q$ is the ``infrared renormalisation group invariant transmutation scale" induced at the quantum level by the radiative corrections of the soft supersymmetry breaking terms at low energies\cite{Noscale,NoscaleTSR}. When $T(t)\le  Q$, the electroweak phase transition takes place,  $SU(2)\times U(1)\to U(1)_{\rm em}$. This starts to be the case at a time $t_W$ and,  for $t>t_W$, the supersymmetry breaking scale $M$ is stabilized at a value close to $Q$. In earlier cosmological times where $M(t),T(t)\gg Q$, the transmutation scale is irrelevant and does not modify our analysis.
\red{An important consequence of this scenario, is a dynamical explanation of the supersymmetry breaking scale.  Indeed, extrapolating the $RDS^{d=4}$ up to the low energy regime where $T={\rm \cal O}(1~TeV)$  one finds, (thanks to the attractor mechanism),  that the natural value of the supersymmetry breaking scale $M(t)$ is naturally small and around the electroweak phase transition, independently of its initial value at early cosmological times.}

\red{ The above statement is absolutely correct, if one assumes that there are no other mass scales created in the infrared, like for instance the various ``infrared renormalization group invariant scales Q" associated with: (i) $\Lambda_{G}$ of hidden gauge group(s), or (ii) the transmutation  scale(s) $Q_H$, (induced in the infrared),  by the renormalized structure of the ``supersymmetry breaking terms".  The existence of non-trivial dynamical scale(s) $Q$ modify the very late cosmological evolution, namely after the electroweak phase transition around $T = T_W \sim Q_H \sim {\rm \cal O}(1TeV)$. Due to this, the intermediate cosmological regime which we study in this paper is defined by : $ T_E \gg T(t), M(t) \gg T_W$.}

Obviously, the physics for $t\gg t_W$ is of main importance in (astro)particle physics and late time cosmology. Unfortunately, the infrared phase at $t\gg t_W$ depends strongly on the  specific choice of  four-dimensional $\N_4=1$ superstring vacuum and the way of spontaneously breaking supersymmetry. A lot of work is necessary to determine the initial superstring vacua that leads in late-times to the precise structure of our Universe. On the other hand, we would like  to stress here that the qualitative infrared behavior of the effective stringy ``no-scale" field theory\cite{NoscaleTSR,StringyNoscale} strongly suggests  that we are in an interesting  string evolutionary  scenario after the ``Hagedorn-transition exit", $t>t_E$, connecting cosmology to particle physics.

\red{Secondly, in the regime close to or at the Hagendorn transition,} the conventional notions of General Relativity such as geometry and topology are well defined only in the low energy and/or small curvature approximations of
a string theory setup \cite{CosmoTopologyChange}. In the very early times of the Universe, $t<t_E$,  purely stringy phenomena at very small  distances and strong curvature scales  imply that the physics   could be quite different from what
one might expect from a ``naive" field theory point of view \cite{CosmoTopologyChange}. In this early epoch,
classical gravity is no longer valid and has to be replaced by
a more fundamental singularity-free theory such as (super-)string theory\cite{MSDS}. Thus, the main obstruction in such a stringy cosmological framework is the Hagedorn temperature limitation $T<T_H$.  Actually, this is not a pathology but rather the signal that a phase transition from a previous vacuum is taking place. The Hagedorn-like singularities have to be resolved either by a stringy phase transition \cite{AtickWitten,AKADK,GV,BV} or by choosing Hagedorn-free string vacua in the
early stage of the universe\cite{MSDS, GravFluxes}.

In this work, we have bypassed the Hagedorn transition ambiguities by considering arbitrary initial boundary conditions (IBC) at $t_E$, the ``Hagedorn transition exit time".  Thanks to the attraction to ``Radiation-like Universes" in late times, most of the ambiguities are washed out. It is however of fundamental interest to investigate further the early non-geometric era of our universe and to show that the induced IBC at $t_E$ solve naturally the ``flatness'' and ``entropy" problem in late cosmological times. This stringy scenario would be an alternative (or at least complementary) to the conventional inflationary scenarios proposed in field theory.


\section*{Acknowledgements}

We are grateful to R. Brandenberger and N. Toumbas for useful discussions.  H.P. thanks the Ecole Normale Sup\'erieure for hospitality.\\
\noindent This work is partially supported by the ANR (CNRS-USAR) contract 05-BLAN-0079-02. The work of F.B., J.E. and H.P. is also supported by the European ERC Advanced Grant 226371 and CNRS PICS contracts  3747 and 4172. J.E. acknowledges financial support from the Groupement d'Int\'er\^et Scientifique P2I.

\newpage
\appendix

\begin{center}{\bf \Large Appendix}\end{center}

\section{Partition function}
\renewcommand{\theequation}{A.\arabic{equation}}
\setcounter{equation}{0}

\subsection{Heterotic string on tori}

We first focus on the heterotic string compactified on a Euclidean toroidal space, while heterotic and type IIB orbifolds will be considered in appendix \ref{Zorbi}. Our starting point is the heterotic string in a background $S^1_E \times T^{D} \times T^n$ ($n=9-D$), where $S^1_E(R_0)$ is the Euclidean time compactified on a circle of radius $R_0$. For simplicity, we choose the tori $T^D$  and $T^{9-D}$ to be products of circles $\Pi_{p=1}^D S^1(R_p)$ and $\Pi_{i=D+1}^9 S^1(R_{i})$. The partition function vanishes, due to supersymmetry:
\be
\label{partAp}
Z = \int_F {d \tau_1 d \tau_2 \over 2 \tau_2} {1 \over 2} \sum_{a,b} (-1)^{a+b+a b} \vartheta^4 \ab \, {\Gamma_{(0,16)} \over \eta^{12} \bar \eta^{24}} \times\Gamma_{(1,1)} (R_0) \times \prod_{p=1}^D \Gamma_{(1,1)}(R_p)  \times \prod_{i=D+1}^9 \Gamma_{(1,1)}(R_i) .
\ee
To implement finite temperature, we deform the $\Gamma_{(1,1)}(R_0)$ lattice by coupling the space-time fermion number $Q_F\equiv a$ to the momentum along $S^1_E(R_0)$.
We also introduce a spontaneous supersymmetry breaking by coupling R-symmetry charges $a+Q_i$  ($i=D+1,\dots,9$) to the momenta along the $T^n$ directions. Spin statistics and modular invariance then determines the precise replacement of the lattices as
\be
\label{phases}
\forall i\in I_b=\{0,D+1,\dots,9\},\, \Gamma_{(1,1)} (R_i) \rightarrow {R_i \over \sqrt{\tau_2}}  \sum_{\tilde m,n} e^{-\pi {R_i^2\over \tau_2} |\tilde m_i+ n_i \tau|^2 } (-1)^{\tilde m_i (a+Q_i) + n_i (b+L_i) + \varepsilon_i\tilde m_i n_i}.
\ee
$I_b$ is the set of labels associated to directions with fluxes that break spontaneously supersymmetry.  In practice, the $Q_i$'s ($i\in I_b$) are linear combinations of charges of the $E_8\times E_8'$ lattice, for which $\varepsilon_i$ is determined to be $0$ or $1$. In our notations, $Q_0=L_0\equiv 0$ and $\varepsilon_0=1$.
A convenient rewriting of the phases in Eq. (\ref{phases}) is done by defining $\tilde m_i=2\tilde k_i+\tilde g_i$, $n_i=2l_i +h_i$ and summing over $\tilde g_i, h_i \in \{0,1\}$ and $\tilde k_i, l_i$ over all integers.
We may evaluate the sum over the spin structures $a$ and $b$ by redefining $a=\hat a + \sum_{i\in I_b} \tilde h_i$ and $b=\hat b + \sum_{i\in I_b} \tilde g_i$. The phases from (\ref{partAp}) and (\ref{phases}) combine to give
\be
a + b + a b + \sum_{i\in I_b} \left( \tilde g_i (a + Q_i) + h_i (b + L_i) + \varepsilon_i \tilde g_i \tilde h_i \right) =\hat a + \hat b + \hat a \hat b + \sum_{i\in I_b} \tilde g_i (1 + Q_i) +  P(\vec {\tilde g}, \vec h, \vec Q,\vec L,\vec \varepsilon)
\ee
where $P(\vec {\tilde g}, \vec h, \vec Q,\vec L,\vec \varepsilon)$ consists of terms which vanish when $h_i=0$ ($\forall i\in I_b$).
We may now make use of the Jacobi identity
\be
\half \sum_{\hat a,\hat b} (-1)^{\hat a+\hat b+\hat a \hat b} \vartheta^4 [{}^{\hat a+\sum_{i} h_i}_{\hat b+\sum_i \tilde g_i}] = - \vartheta^4 [{}^{1+\sum_{i} h_i}_{1+\sum_{i} \tilde g_i}]
\ee
to obtain the result
\bea
\label{parti2}
Z\!\!\! &=& \!\!\! -\int_F {d \tau_1 d \tau_2 \over 2 \tau_2} \sum_{ \tilde g_j, h_j  (j\in I_b)} (-1)^{\sum_k \tilde g_k (1 + Q_k) } (-1)^{P(\vec {\tilde g}, \vec h, \vec Q,\vec L,\vec \varepsilon)}
\vartheta^4[{}^{1+\sum_{q}  h_q}_{1+\sum_{q} \tilde g_q}]
\no\\&& \hskip0.4in
\prod_{i\in I_b} \Gamma_{(1,1)}[{}^{h_i}_{\tilde g_i}](R_i) \times \prod_{p=1}^{D} \Gamma_{(1,1)}(R_{p})
\times {\Gamma_{(0,16)} \over \eta^{12} \bar \eta^{24}},
\eea
where we have introduced the shifted lattices
\be
\label{shift}
\Gamma_{(1,1)}[{}^{h_i}_{\tilde g_i}](R_i) = {R_i\over \sqrt{\tau_2}} \sum_{\tilde k_i,l_i} e^{- \pi {R_i^2\over \tau_2} |2 \tilde k_i + \tilde g_i + (2  l_i + h_i) \tau|^2}.
\ee

Important simplifications can be made, using the fact that $R_i\gg 1$ ($i\in I_b$):

\noindent - When $2l_i+h_i\neq 0$ in Eq. (\ref{shift}), the integrand in the partition function (\ref{parti2}) contains a factor $e^{-\pi R_i^2(2l_i+h_i)\tau_2}$ implying an exponentially suppressed contribution after integration. We thus restrict to the sectors with $l_i=h_i=0$ ($i\in I_b$).

\noindent - The sectors $\sum_i h_i=\sum_i \tilde g_i=0$ mod 2 are supersymmetric and therefore do not contribute, as can be seen from the presence of a $\vartheta^4[^1_1]$ factor in (\ref{parti2}). We thus only keep the sectors $h_i=0$ ($i\in I_b$), $\sum_i \tilde g_i=1$ mod 2.

\noindent - All of them, have at least one $i\in I_b$ such that $\tg_{i}=1$, so that the integrand in (\ref{parti2}) contains a factor $e^{-\pi {R_{i}^2\over \tau_2}}$. This implies that we can extend the integral over the fundamental domain, to an integral over the full upper half strip. The error introduced this way is exponentially suppressed\footnote{Note that if the model before introducing temperature and internal fluxes was not supersymmetric, the sector $h_i=\tg_i=0$ ($\forall i\in I_b$) would not vanish. Its integral over the fundamental domain would not be suppressed by any $e^{-\pi {R_{i}^2\over \tau_2}}$ ($i\in I_b$) factor and could not be replaced by the integral over the strip. The result would imply a very large (but finite) contribution to the vacuum energy, proportional to the number of massless bosons minus the number of massless fermions.}.

\noindent Altogether, the partition function reduces to
\be
\label{part2}
\begin {array}{ll}
\dis Z = \left( \prod_{k\in I_b}R_k\right) \int_{-\half}^\half d\tau_1\int_0^{+\infty} {d \tau_2 \over 2 \tau_2^{n+3\over 2}} &\!\!\!\!\!\!\!\dis\sum_{ \scriptsize \begin{array}{c}\tk_j, \tilde g_j (j\in I_b)\\ \sum_{j} \tilde g_j = 1 \rm mod \, 2\end{array}} e^{-{\pi\over \tau_2}\sum_i R_i^2 (2\tk_i+\tg_i)^2}\\
&\dis\times  (-1)^{\sum_k \tilde g_k Q_k } {\vartheta^4[{}^{1}_{0}] \over \eta^{12} \bar \eta^{24}} \Gamma_{(0,16)} \times \prod_{p=1}^{D} \Gamma_{(1,1)}(R_{p}).
\end{array}
\ee

In this expression, the low lying contributions from the oscillators and right moving lattice $\Gamma_{(0,16)}$ are
\be
\label{internalsector}
(-1)^{\sum_{k\in I_b} \tilde g_k Q_k } {\vartheta^4[{}^{1}_{0}] \over \eta^{12} \bar \eta^{24}} \Gamma_{(0,16)} = 2^4 \left( {1\over \bar q} + D_0(\vec\tg,\vec Q) + \cO(q, \bar q) \right),
\ee
where $q = e^{2 i\pi  \tau}$ and $D_0(\vec \tg,\vec Q)$ is the sum over massless degrees of freedom with each mode weighted by the factor $(-1)^{\sum_k \tg_k Q_k}$. Defining $I_s=\{1,\dots,D\}$ the set of ``spectator'' directions \ie that do not break supersymmetry, the lattice of $T^D$ can also be expanded as
\be
\label{spec}
\prod_{p=1}^D \Gamma_{(1,1)} (R_p) =  \sum_{ m_q, n_q (q\in I_s)}e^{-\pi\tau_2\sum_p\left( ({m_p\over R_p})^2+(n_p R_p)^2\right)}e^{-2i\pi\tau_1\sum_p m_pn_p}.
\ee
The change of variable $\tau_2=x\pi \left( \sum_{i\in I_b}R_i^2(2\tk_i+\tg_i)\right)$ in the $\tau_2$-integration shows that the massive contributions in Eqs (\ref{internalsector}) and (\ref{spec}) are exponentially suppressed, compared to the massless ones. We thus concentrate our attention on the light degrees of freedom.
Because we have been able to replace the fundamental domain with the half strip, the integration over $\tau_1$ now simply enforces the level matching condition. The constant term in the r.h.s. of Eq. (\ref{internalsector}) combined with Eq. (\ref{spec}) implies that it is enough to keep for each $p\in I_s$,
\be
\label{condi}
\left\{\begin{array}{ll}
n_p=0 & \dis \mbox{if } R_p\gg \max_{i\in I_b}R_i,\\
m_p=0 & \dis\mbox{if } R_p\ll { 1\over \dis \max_{i\in I_b}R_i},\\
m_p=n_p=0 &\dis \mbox{else.} \\
\end{array}\right.
\ee
Similarly, the $\bar q^{-1}$ term in Eq. (\ref{internalsector}) combined with Eq. (\ref{spec}) implies that it is enough to keep  for each $q\in I_s$,
\be
\label{2Higgs}
m_q=n_q=\pm 1\quad \mbox{if }R_q\simeq 1 \quad \mbox{with } \quad  m_p, n_p \mbox{ given in Eq. (\ref{condi}), }\forall p\in I_s, p\neq q .
\ee
Physically, the above two winding modes are responsible for the gauge symmetry enhancement $U(1)\to SU(2)$ at $R_q=1$.  Away from the self dual point, they are massive: Their contribution to $Z$ becomes exponentially negligible, while the $SU(2)$ is Higgsed. The enhancement of the symmetry will play an important role in stabilizing $R_q$ around one.
The following reduced expression for the partition function when $R_p\ge 1$ ($\forall p\in I_s$) is then obtained,
\be
\label{Z2}
\begin {array}{ll}
\dis Z = &\!\!\!\dis\left( \prod_{k\in I_b}R_k\right) {2^4\over 2}\sum_{ \scriptsize \begin{array}{c}\tk_j, \tilde g_j (j\in I_b)\\ \sum_{j} \tilde g_j = 1 \rm mod \, 2\end{array}} {2^4\over 2}\int_0^{+\infty} {d \tau_2 \over  \tau_2^{n+3\over 2}} \exp\left({-{\pi\over \tau_2}\sum_{i\in I_b} R_i^2 (2\tk_i+\tg_i)^2}\right)\\
& \!\!\!\!\! \dis \times\Bigg\{D_0(\vec \tg,\vec Q)\sum_{m_r (r\in I_s)}\exp\left(-\pi\tau_2\sum_{p\in I_s}\left({m_p\over R_p}\right)^2\right)\\
& \!\!\!\!\!  \dis + 2 \sum_{q\in I_s}\exp\left(-\pi\tau_2\left({1\over R^2_q}+R_q^2-2\right)\right)\sum_{m_r (r\in I_s, r\neq q)}\exp\left(-\pi\tau_2\sum_{p\in I_s, p\neq q}\left({m_p\over R_p}\right)^2\right)\Bigg\}.
\end{array}
\ee
The companion expressions when any $R_p\le 1$ ($p\in I_s$) is obtained by replacing $\dis R_p\to {1\over R_p}$.

The above result can be rewritten in various forms. We introduce an integer parameter $\Delta$ and split the $T^{D}$ torus as $T^{D-\Delta} \times T^\Delta$. Correspondingly, the ``spectator'' indices $I_s$ are divided in two sets,
\be
I_s^l=\{1,\dots,D-\Delta\}\; , \qquad I_s^s=\{D-\Delta+1,\dots,D\},
\ee
and we perform a Poisson resummation on the $T^{D-\Delta}$ zero mode lattice. The reason for that is that when some $R_p$ ($p\in I_s$) is ``small'' \ie $R_p\ll \inf_{i\in I_b}R_i$ , the Hamiltonian formulation of Eq. (\ref{Z2}) is relevant, while when some $R_p$  ($p\in I_s$) is ``large'' \ie $R_p\gg \max_{i\in I_b}R_i$, the Lagrangian formulation is more convenient. The alternative forms for arbitrary $\Delta$ are:
\be
\begin {array}{ll}
\dis Z = &\!\!\!\dis\left( \prod_{k\in I_b}R_k\right) \left( \prod_{s'\in I_s^l}R_{s'}\right)\sum_{ \scriptsize \begin{array}{c}\tk_j, \tilde g_j (j\in I_b)\\ \sum_{j} \tilde g_j = 1 \rm mod \, 2\end{array}} {2^4\over 2}\int_0^{+\infty} {d \tau_2 \over  \tau_2^{D-\Delta +n+3\over 2}} \exp\left(-{\pi\over \tau_2}\sum_{i\in I_b} R_i^2 (2\tk_i+\tg_i)^2\right)\\
& \dis \times\Bigg\{D_0(\vec \tg,\vec Q)\sum_{\scriptsize\begin{array}{ll}\tm_{r'} (r'\in I^l_s)\\m_r (r\in l_s^s)\end{array}}
\exp\left(-{\pi\over \tau_2}\sum_{p'\in I_s^l}(\tm_{p'}R_{p'})^2-\pi\tau_2\sum_{p\in I_s^s}\left({m_p\over R_p}\right)^2\right)\\
& \quad  \dis + 2 \sum_{q\in I_s^s}\exp\left(-\pi\tau_2\left({1\over R^2_q}+R_q^2-2\right)\right)\\
&\quad \qquad \dis \sum_{\scriptsize \begin{array}{l}\tm_{r'} (r'\in I_s^l)\\m_r (r\in l_s^s, r\neq q)\end{array}}
\exp\left(-{\pi\over \tau_2}\sum_{p'\in I_s^l}(\tm_{p'}R_{p'})^2-\pi\tau_2\sum_{p\in I_s^s, p\neq q}\left({m_p\over R_p}\right)^2\right)\\ \nonumber
\end{array}
\ee
\be
\begin {array}{ll}
\
& \quad  \dis + 2 \sum_{q'\in I_s^s}{\sqrt{\tau_2}\over R_{q'}}\exp\left(-\pi\tau_2\left({1\over R^2_{q'}}+R_{q'}^2-2\right)\right)\\
&\quad \qquad \dis \sum_{\scriptsize \begin{array}{l}\tm_{r'} (r'\in I_s^l, r'\neq q')\\m_r (r\in l_s^s)\end{array}}
\exp\left(-{\pi\over \tau_2}\sum_{p'\in I_s^l, p'\neq q'}(\tm_{p'}R_{p'})^2-\pi\tau_2\sum_{p\in I_s^s}\left({m_p\over R_p}\right)^2\right)\Bigg\}.
\end{array}
\ee

Using the integral form of the modified Bessel function of the second kind $K_\alpha(z)$,
\be
\int_{0}^{+\infty} {d \tau_2 \over \tau_2^\alpha} \exp \left( - {\pi\over \tau_2} F \right) \exp (- \pi  \tau_2 G)= \left({G\over F}\right)^{(\alpha - 1) \over 2} 2 K_{\alpha-1}(2 \pi \sqrt{F G}),
\ee
one obtains
\be
\label{part2het}
\begin {array}{ll}
\dis Z = &\dis\left( \prod_{k\in I_b}R_k\right) \left( \prod_{s'\in I_s^l}R_{s'}\right) {2^4\over 2}\sum_{\scriptsize \begin{array}{c}\tk_j, \tilde g_j (j\in I_b)\\ \sum_{j} \tilde g_j = 1 \rm mod \, 2\end{array}}\\
& \dis \times\Bigg\{D_0(\vec \tg,\vec Q)\sum_{\tm_{r'} (r'\in I^l_s)}\Bigg[{\Gamma\left({D-\Delta+n+1\over 2}\right)\over (\pi F_1)^{D-\Delta+n+1\over 2}}+{\sum_{m_r (r\in l_s^s)}}^{\!\!\!\!\prime\,\,\,}\left({G_1\over F_1}\right)^{D-\Delta+n+1\over 4}2K_{D-\Delta+n+1\over 2}(2\pi\sqrt{F_1G_1})\Bigg]\\
& \dis \quad  +2\sum_{q\in I_s^s}\sum_{\scriptsize \begin{array}{l}\tm_{r'} (r'\in I_s^l)\\m_r (r\in l_s^s, r\neq q)\end{array}} \left({G_2\over F_1}\right)^{D-\Delta+n+1\over 4}2K_{D-\Delta+n+1\over 2}(2\pi\sqrt{F_1G_2})\\
&\quad  \dis  +2\sum_{q'\in I_s^l}\sum_{\scriptsize \begin{array}{l}\tm_{r'} (r'\in I_s^l,r'\neq q')\\m_r (r\in l_s^s)\end{array}}{1\over R_{q'}} \left({G_3\over F_3}\right)^{D-\Delta+n\over 4}2K_{D-\Delta+n\over 2}(2\pi\sqrt{F_3G_3})\Bigg\},
\end{array}
\ee
where we have defined
\be
\label{FG}
\begin{array}{ll}
\!\!\!\!\dis F_1=\sum_{i\in I_b} R_i^2 (2\tk_i+\tg_i)^2+ \sum_{p'\in I_s^l}(\tm_{p'}R_{p'})^2, &\dis  G_1=\sum_{p\in I_s^s}\left({m_p\over R_p}\right)^2,\\
\!\!\!\!&\dis G_2=\left({1\over R_q}-R_q\right)^2+\sum_{p\in I_s^s,p\neq q}\left({m_p\over R_p}\right)^2,\\
\!\!\!\!\dis F_3=\sum_{i\in I_b} R_i^2 (2\tk_i+\tg_i)^2+ \sum_{p'\in I_s^l,p'\neq q'}(\tm_{p'}R_{p'})^2, &\dis  G_3=\left({1\over R_{q'}}-R_{q'}\right)^2+\sum_{p\in I_s^s}\left({m_p\over R_p}\right)^2.
\end{array}
\ee
In the second line of Eq. (\ref{part2het}), the ``primed'' sum in the brackets means that $m_r=0$ ($\forall r\in I_s^s$) is excluded.
We remind the reader that Eq. (\ref{part2het}) is valid when  $R_p \geq 1$ ($\forall p\in l_s^l\cup l_s^s$).  The expressions with some  $R_q$'s such that $R_q\leq 1$ is obtained by T-duality  \ie by exchanging them with their inverses, $R_q\to \dis{1 \over R_q}$, in Eqs (\ref{part2het}) and (\ref{FG}).

It will be convenient to have the rules for decompactifying directions of $T^{D-\Delta}$ as well as the rules for freezing one of the $T^\Delta$ radii.  \begin{itemize}
\item For decompactifying a radius $R_{s'}$ ($s' \in I^l_s$), one simply keeps only the terms with $\tilde m_{s'} = 0$ in (\ref{part2het}).  In the last line, one also discards the term with $q' = s'$.  The net result is a remaining overall factor of $R_{s'}$ in the first line.
\item In order to freeze a radius $R_s$ ($s\in I_s^s$) at the self dual point $R_s=1$, one keeps only the terms with $m_s = 0$ in (\ref{part2het}).  In addition, one discards the term with $q = s$ in the third line and shifts $D_0 \rightarrow D_0 + 2$.
\item In order to freeze a radius $R_s$ ($s\in I_s^s$) at an arbitrary value such that $1/R_s$ and $R_s\ll \inf_{i\in I_b}R_i$, one keeps only the terms with $m_s = 0$ in (\ref{part2het}).  In addition, one discards the term with $q = s$ in the third line.
\end{itemize}

\subsection{Heterotic and type IIB orbifolds}
\label{Zorbi}

We next consider the heterotic string on $\dis S^1_E \times T^D \times {T^4 \over  \Z_2}\times T^n $ ($n=5-D$). The partition function vanishes, due to the 8 conserved space-time supercharges,
\be
\begin{array}{ll}
\dis Z = \int_F {d \tau_1 d \tau_2 \over 2 \tau_2}& \!\!\!\dis {1 \over 2} \sum_{H,G}{1 \over 2} \sum_{a,b} (-1)^{a+b+a b} \vartheta^2 \ab \vartheta[^{a+H}_{b+G}]\vartheta[^{a-H}_{b-G}]\, Z_{(4,4)}[^H_G]\, {\Gamma_{(0,16)}[^H_G] \over \eta^{8} \bar \eta^{20}}\\
&\dis  \times\Gamma_{(1,1)} (R_0) \times \prod_{p=1}^D \Gamma_{(1,1)}(R_p)  \times \prod_{i=D+5}^9 \Gamma_{(1,1)}(R_i),
\end{array}
\ee
where $H, G$ are equal to 0 or 1. $Z_{(4,4)}[^H_G]$ is the block that accounts for the $\dis {T^4 \over  \Z_2}$ part of the background and the right moving $\Gamma_{(0,16)}[^H_G]$ lattice is consistently $\Z_2$-twisted to guarantee modular invariance, thereby breaking part of the initial gauge group.  Temperature and supersymmetry breaking are again introduced by modifying the lattice sums along $S^1_E$ and $T^n$ as in (\ref{phases}). However, the left moving charges $Q_i$ may now also involve the ``orbifold twist number'' $H$.

The analysis of the toroidal case can be applied the same way. Defining $I_b=\{0,D+5,\dots,9\}$, any sector with some $h_i \neq 0$ ($i\in I_b$) is exponentially suppressed and the sectors with $\sum_{i\in I_b} h_i = 0$ and $\sum_{i\in I_b} \tilde g_i = 0$ vanish due to supersymmetry.
We may again replace the fundamental domain with the full upper half strip.  We treat $\dis {T^4 \over  \Z_2}$ as part of the internal sector, with frozen moduli much smaller than $\inf_{i\in I_b}R_i$. As before, the non-exponentially suppressed contributions to the partition function arise from the massless modes (and their light towers of KK (or winding) states).
For explicit examples, see \cite{Cosmo-1}.  The net result is that the partition function is of the same form as in (\ref{part2het}) except that the numbers $D_0(\vec\tg,\vec Q)$ are different.

The type IIB partition function on $\dis S^1_E \times T^D \times {T^4 \over  \Z_2} \times T^n$ ($n=5-D$)  takes the form
\be
\label{partIIBa}
\begin{array}{ll}
\dis Z = \int_F {d \tau_1 d \tau_2 \over 2 \tau_2}& \!\!\!\dis {1\over 2}\sum_{H,G} {1 \over 2 } \sum_{a,b} (-1)^{a+b+a b} \vartheta^2 \ab \vartheta[^{a+H}_{b+G}]\vartheta[^{a-H}_{b-G}]
\, {1 \over 2} \sum_{\bar a,\bar b} (-1)^{\bar a+\bar b+\bar a \bar b} \bar \vartheta^2 \abb \bar \vartheta[^{\bar a+H}_{\bar b+G}]\bar \vartheta[^{\bar a-H}_{\bar b-G}]\\
& \!\!\!\dis\times {Z_{(4,4)}\over \eta^{12}\bar\eta^{12}} \times \Gamma_{(1,1)} (R_0) \times \prod_{p=1}^D \Gamma_{(1,1)}(R_p)  \times \prod_{i=D+5}^9 \Gamma_{(1,1)}(R_i).
\end{array}
\ee
Temperature and supersymmetry breaking are again introduced by modifying the lattice sums along $S_E^1$ and $T^n$. However, in the type IIB case, we may introduce phases similar to Eq. (\ref{phases}) but involving either left moving R-charges $a+Q_i$ ($i\in I_b$), or right moving ones $\bar a+\bar Q_i$, or both.  In the present paper, we consider cases where both left and right charges are non-trivial. Some models in this class where shown to allow critical cosmological evolutions corresponding to radiation\red{-like} eras \cite{Cosmo-2, Francois}.

For $R_i\gg 1$ ($\forall i\in I_b$), the manipulations used in the heterotic case can be applied similarly, up to an important difference. In the sector $[^H_G]=[^0_0]$, the analogue of the heterotic contribution given in Eq. (\ref{internalsector}) is in type IIB,
\be
(-1)^{\sum_{k\in I_b} \tg_k (Q_k + \bar Q_k)} {\vartheta^4[{}^{1}_{0}] \bar \vartheta^4[{}^{1}_{0}] \over \eta^{12} \bar \eta^{12}} = 2^4 \bigg(1 + \cO(q, \bar q) \bigg).
\ee
Consequently, there is no massless winding mode arising when some $R_q=1$ ($q\in I_s$), as opposed to the heterotic states given in (\ref{2Higgs}).  The final form of the partition function is then formally identical to the two first lines of the heterotic one (\ref{part2het}), with coefficient $D_0(\vec \tg,\vec Q,\vec{\bar Q})$,
\be
\label{part2ii}
\begin {array}{ll}
\dis Z = &\!\!\!\dis\left( \prod_{k\in I_b}R_k\right) \left( \prod_{s'\in I_s^l}R_{s'}\right) {2^4\over 2}\sum_{\scriptsize \begin{array}{c}\tk_j, \tilde g_j (j\in I_b)\\ \sum_{j} \tilde g_j = 1 \rm mod \, 2\end{array}}\\
& \dis \!\!\!\times\Bigg\{D_0(\vec \tg,\vec Q,\vec{\bar Q})\sum_{\tm_{r'} (r'\in I^l_s)}\Bigg[{\Gamma\left({D-\Delta+n+1\over 2}\right)\over (\pi F_1)^{D-\Delta+n+1\over 2}}+{\sum_{m_r (r\in l_s^s)}}^{\!\!\!\!\prime\,\,\,}\left({G_1\over F_1}\right)^{D-\Delta+n+1\over 4}2K_{D-\Delta+n+1\over 2}(2\pi\sqrt{F_1G_1})\Bigg],
\end{array}
\ee
where $F_1$, $G_1$ are defined in Eq. (\ref{FG}).

\section{Equations of motion}
\renewcommand{\theequation}{B.\arabic{equation}}
\setcounter{equation}{0}
\label{sececmo}

In this appendix, we derive the equations of motion in our thermal backgrounds. The 9 space-like directions are $T^D\times T^n$ in the toroidal models and $T^D\times \dis {T^4\over \Z_2}\times T^n$ in the orbifold ones. To unify the two cases, we define $n=9-A-D$, where $A=0$ in toroidal compactifications and $A=4$ in orbifold ones. Geometrical fluxes in the $T^n$ torus break spontaneously supersymmetry. A $T^{d-1}$ sub-torus of $T^D$ is taken to be  isotropic and very large, to be interpreted as part of the spatial directions of the space-time. In practice, this means that the radii of $T^{d-1}$ are proportional for all time $t$ and that the associated KK states are continuous, implying the convergence of the partition function.  The remaining $T^{D-d+1}$ sub-torus is allowed to have arbitrary radii.  When they are all small, we interpret them as internal and the space-time dimension is $d$.  When only $\Delta$  of the $T^{D-d+1}$ radii ($D-(d-1) \geq \Delta \geq 0$) are small, the space-time dimension is then ${\cal D} \equiv D  + 1 - \Delta$. However, this space-time with enhanced dimension  is anisotropic since only part of its space-like radii are evolving proportionally.

Our starting point is the standard $10$-dimensional dilaton-gravity theory,
\be
S=\half \int d^{10} x \sqrt{-\hat g_{10}}\,  e^{-2\dil {}_{10}}\left[\hat R_{10} + 4 \partial_\mu \dil {}_{10} \partial^\mu \dil {}_{10} \right] - \int d^{10} x \sqrt{-g_{10}} \, {\hat {\cal F}}_{10}.
\ee
The hats denote that we are in string frame and the numerical subscripts, here 10,  indicate the space-time dimension. $\dil {}_{10}$ and $\hat {\cal F}_{10}$ are the dilaton and free energy density in 10 dimensions. The latter is related to the partition function by ${\hat {\cal F}}_{10} = \dis -{Z \over V_{10}}$, where $V_{10}$ is the 10-dimensional volume of the Euclidean background in which we computed $Z$.  We split the $T^{D}$ torus as $T^{{\cal D}-1} \times T^\Delta$ and dimensionally reduce on $T^\Delta$ and $T^n$ (or $\dis  {T^4\over \Z_2}\times T^n$ in the orbifold models).  The action becomes
\bea
\label{S1}
S\!\!\! &=& \!\!\!\half \int d^{{\cal D}} x \sqrt{-\hat g_{{\cal D}}}\,  e^{-2\dil {}_{{\cal D}}}\left[\hat R_{{\cal D}} + 4 \partial_\mu \dil {}_{\cal D} \partial^\mu \dil {}_{\cal D} - \sum_{p=D-\Delta+1}^{D} {\partial_\mu R_p \partial^\mu R_p \over R_p^2} - \sum_{i=D+1}^9 {\p_\mu R_i \p^\mu  R_i \over  R_i^2}
\right] \no\\&& \hskip1.0in - \int d^{{\cal D}} x \sqrt{-\hat g_{{\cal D}}}\,  {\hat {\cal F}}_{{\cal D}},
\eea
where the dilaton in ${\cal D}$-dimensions is $\dil {}_{\cal D} = \dil {}_{10} - \half \sum_{\gamma=D-\Delta+1}^9 \ln 2 \pi R_\gamma$ and ${\hat {\cal F}}_{\D} = \dis-{Z \over  V_{\D}}$. In some instances, we will suppose for simplicity that some internal radii are frozen. This will be the case for some radii of $T^\Delta$, and the radii of $\dis {T^4\over \Z_2}$ in the orbifold models. Then, the corresponding kinetic terms would simply disappear from the action (\ref{S1}). Performing the conformal transformation $g_{\cal D} =  \exp ( -{4 \over {\cal D} - 2} \dil {}_{\cal D} ) \hat g_{\cal D}$ brings us to Einstein frame and the action becomes
\bea
S \!\!\!&=& \!\!\!\half \int d^{\cal D} x \sqrt{-g_{\cal D}} \left[R_{{\cal D}} - {4 \over {\cal D} - 2} \partial_\mu \dil {}_{{\cal D}} \partial^\mu \dil {}_{{\cal D}} - \sum_{p=D-\Delta+1}^{D} {\partial_\mu R_p \partial^\mu R_p \over R_p^2} - \sum_{i=D+1}^9 {\p_\mu R_i \p^\mu R_i \over R_i^2}
\right] \no\\&& \hskip1.0in - \int d^{{\cal D}} x \sqrt{-g_{{\cal D}}}\,  {\cal F}_{{\cal D}},
\eea
where ${\cal F}_{{\cal D}} = \exp ( {2 {\cal D} \over {\cal D} - 2} \dil {}_{{\cal D}} ) {\hat {\cal F}}_{{\cal D}}$.  The supersymmetry breaking scale measured in Einstein frame, $M_{\cal D}$, is given by the inverse volume of $T^n$,
\bea
M_{\cal D} = e^{ 2 \dil {}_{{\cal D}} \over {\cal D} - 2}  \prod_{i = D+A+1}^{9} {1\over (2\pi R_i)^{{1 /n}}} = {1\over 2\pi}\, \exp \bigg( {{2 \dil {}_{{\cal D}} \over {\cal D} - 2} - {1 \over n} \sum_{i = D+A+1}^{9} \ln R_i} \bigg).
\eea
It is convenient to introduce an explicit field notation, $\Phi_{\cal D}$, for the supersymmetry breaking scale as
\be
\label{MD}
M_{\cal D} = {e^{\alpha \Phi_{\cal D}} \over 2 \pi}\qquad \where \qquad \alpha^2 = {1 \over {\cal D} - 2} + {1 \over n}.
\ee
The coefficient $\alpha$ is chosen so that $\Phi_{\cal D}$ has a canonically normalized kinetic term. We can introduce other fields, $\phi_{\bot \cal D}$ and $\varphi_i$ ($i = 1,\dots,n-1$), to describe the remaining degrees of freedom among the dilaton and radii of $T^n$.  The explicit transformation law is given as
\bea
\label{orthofi}
\left(
  \begin{array}{c}
    \Phi_{\cal D} \\
    \phi_{\bot {\cal D}} \\
    \varphi_{n-1} \\
    \vdots \\
    \varphi_1 \\
  \end{array}
\right)
=
\left(
  \begin{array}{ccccc}
    {1 \over \alpha \sqrt{{\cal D} -2}} & -{1 \over \alpha n} & -{1 \over \alpha n} & \ldots & -{1 \over \alpha n} \\
    \sqrt{ {\cal D} - 2 \over {\cal D}+ n - 2} & {1 \over \sqrt{{\cal D} + n -2}} & {1 \over \sqrt{{\cal D} + n -2}} & \ldots & {1 \over \sqrt{{\cal D} + n -2}} \\
    0 & {n-1 \over \sqrt{n (n-1)}} & -{1 \over \sqrt{n (n-1)}} & \ldots & -{1 \over \sqrt{n (n-1)}} \\
    \vdots &   &  \ddots   &  & \vdots \\
    0 & \ldots & 0 & - {1 \over \sqrt{2}} & -{1 \over \sqrt{2}} \\
  \end{array}
\right)
\left(
  \begin{array}{c}
    {2 \over \sqrt{ {\cal D} - 2}} \dil {}_{\cal D} \\
    \ln R_{9} \\
    \ln R_{8} \\
   \vdots \\
    \ln R_{D+A+1} \\
  \end{array}
\right).
\eea
Finally, we denote the $\Delta$ degrees of freedom of $T^\Delta$ as $\zeta_p=\ln R_{D-\Delta+p}$ ($p=1, \dots,\Delta$).
In terms of these fields, the action takes the canonical form\footnote{One could also introduce $A=4$ more scalars associated to the radii of $\dis{T^4/\Z_2}$ in the orbifold models. However, as announced before, we consider from now on these radii to be internal and constant.}
\bea
\label{Aaction}
S &=& \int d^{{\cal D}} x \sqrt{-g_{{\cal D}}} \bigg[ {R_{{\cal D}} \over 2} - \half \partial_\mu \phi_{\bot {\cal D}} \partial^\mu \phi_{\bot {\cal D}} - \half \partial_\mu \Phi_\D \partial^\mu \Phi_\D - \half \sum_{i=1}^{n-1} \partial_\mu \varphi_i \partial^\mu \varphi_i- \half \sum_{p=1}^\Delta \p_\mu \zeta_p \p^\mu \zeta_p
\bigg]
\no\\&& \hskip1.0in
 - \int d^{{\cal D}} x \sqrt{-g_{{\cal D}}}\,  {\cal F}_{{\cal D}}.
 \eea

The metric variation gives the standard Einstein equation in terms of the stress energy tensor.  It is convenient to single out the terms coming from the thermal corrections and so we define the thermal energy-momentum tensor as
\bea
\label{TEMT}
{T_{\cal D}}_{\mu \nu} = - {g_{\cal D}}_{\mu \nu} {\cal F}_{{\cal D}} + 2 {\partial {\cal F}_{{\cal D}} \over \partial (g_{{\cal D}}) {}^{\mu \nu}}.
\eea
We are interested in space-times which are homogeneous but anisotropic, since the radii of $T^{{\cal D}-d}$ are allowed to vary independently of the radii of the very large $T^{d-1}$.  In this case, the fields are only allowed to vary with time and we take the metric to be of the form
\be
\label{az}
ds_{\cal D}^2 = - dt^2 + a(t)^2 \sum_{l=1}^{d-1} dx_l^2 + \sum_{k=1}^{{\cal D} - d} b_k(t)^2 dx_{d-1+k}^2.
\ee
The $b_k$ are the scale factors of the $T^{{\cal D}-d}$ torus in Einstein frame and are related to the radii in string frame by $b_k = e^{ {-2 \over {\cal D} - 2} \dil {}_{\cal D}} 2 \pi R_k$ if $R_k\gg 1$ (exchange $R_k\to 1/R_k$ if $R_k\ll 1$).  It is also convenient to introduce the temperature related to the radius of the Euclidean time used in the  computation of $Z$ as
\be
\label{TD}
T_{\cal D} \equiv {e^{ {2 \over {\cal D} - 2} \dil {}_{\cal D}} \over 2 \pi R_0},
\ee
(not to be confused with the stress energy tensor). Defining the thermal energy density $\rho_{\cal D} \equiv {T_{\cal D}}_{00}$ and pressure $P_{\cal D} \equiv a^{-2} {T_{\cal D}}_{ll}$ (no sum on $l=1,\dots,d-1$), the thermal energy-momentum tensor can be expressed as\footnote{The extra temperature factor of $T_{\cal D}^2$ in the second term of the first equation in (\ref{TEMT2}) can be seen as follows.  The variation of the free energy density can be taken when the coordinates are such that the metric is just the analytic continuation of the Euclidean background, $ds_{\cal D}^2 = - T_{\cal D}^{-2} dt^2 + a(t)^2 \sum_l ds_l^2 + \sum_{k} b_k(t)^2 dx_{d-1+k}^2$.  The factor of $T_{\cal D}^2$ then comes from changing the metric coordinates to have (\ref{az}).}
\bea
\label{TEMT2}
{T_{\cal D}}_{00} &=& {\cal F}_{{\cal D}} + 2 T_{\cal D}^{2} {\partial {\cal F}_{{\cal D}} \over \partial T_{\cal D}^2 }
= \bigg( T_{\cal D} {\p P_{\cal D} \over \p T_{\cal D}} - P_{\cal D} \bigg)
\\
\label{p=-F}
{T_{\cal D}}_{ll} &=& - a^2 {\cal F}_{{\cal D}}
= a^2 P_{\cal D} \\
\label{P4}
{T_{\cal D}}_{d+k, d+k} &=& - b_k^2 {\cal F}_{{\cal D}} + 2 {\partial {\cal F}_{{\cal D}} \over \partial b_k^{-2}}
= b_k^2 \bigg( P_{\cal D} + b_k {\p  P_{\cal D} \over \p b_k} \bigg).
\eea
 Note that the thermal energy density $\rho_{\cal D}$ satisfies the thermodynamical identity $\dis \rho_{\cal D} = T_{\cal D} {\p P_{\cal D} \over \p T_{\cal D}} - P_{\cal D}$.
 The Ricci tensor can be expressed in terms of $\dis H \equiv {\dot{a} \over a}$ and $\dis K_k \equiv {\dot{b}_k \over  b_k}$ ($k=1,\dots, \D-d$).  The off-diagonal elements automatically vanish, as well as those of the thermal energy-momentum tensor (\ref{TEMT2}).  The remaining diagonal Einstein equations become
\bea
\label{eomeinstein}
0 &=& (d-1) \bigg( \dot{H} + H^2 \bigg) + \sum_{k=1}^{{\cal D} - d} \bigg( \dot{K}_k + K_k^2 \bigg) +
\dot{\phi}_{\bot {\cal D}}^2 + \dot{\Phi}_{\cal D}^2 + \sum_{i=1}^{n-1} \dot{\varphi}_i^2 + \sum_{i=1}^\Delta {\dot{\zeta}_i}^2
\no\\&& \qquad
+ {{\cal D} - 3 \over {\cal D} - 2} \rho_{\cal D} + {d - 1 \over {\cal D} - 2} P_{\cal D} + {1 \over {\cal D} - 2} \sum_{k=1}^{{\cal D} - d} \bigg( P_{\cal D} + b_k {\p  P_{\cal D} \over \p b_k} \bigg)
\\ && \no\\
\label{eomeinstein2}0 &=& (d-1) H^2 + \dot{H} + H \sum_{k=1}^{\D-d} K_k
- {1 \over {\cal D} - 2} \rho_{\cal D} - {{\cal D} - d - 1 \over {\cal D} - 2} P_{\cal D}
\no\\&&
\qquad + {1 \over {\cal D} - 2} \sum_{k=1}^{{\cal D} - d} \bigg( P_{\cal D} + b_k {\p  P_{\cal D} \over \p b_k} \bigg)
\\ && \no\\
0 &=&
\dot{K}_k + K_k^2 + (d-1) K_k {H} + K_k \sum_{j =1(j\neq k)}^{\D-d} K_j
-{1 \over {\cal D} -2} \rho_{\cal D} - {{\cal D} - 3 \over {\cal D} - 2} \bigg( P_{\cal D} + b_k {\p  P_{\cal D} \over \p b_k} \bigg)
\no\\
\label{eomeinstein3}
&& \qquad
+ {d - 1 \over {\cal D} - 2} P_{\cal D} + {1 \over {\cal D} - 2} \sum_{j=1( j \neq k)}^{{\cal D} - d} \bigg( P_{\cal D} + b_j {\p  P_{\cal D} \over \p b_j} \bigg)\qquad (k=1,\dots,\D-d),
\eea
where ${\cal D} = D + 1 - \Delta$.  The equations of motion for the scalars reduce to
\bea
\label{eomscalars}
&&
\ddot{\Phi}_{\cal D} + (d-1) H \dot{\Phi}_{\cal D} + \sum_{k=1}^{{\cal D} - d} K_k \dot{\Phi}_{\cal D} = {\p P_{\cal D} \over \p \Phi_{\cal D}}
\\&&
\label{eomscalars2}
\ddot{\phi}_{\bot {\cal D}} + (d-1) H \dot{\phi}_{\bot {\cal D}} + \sum_{k=1}^{{\cal D} - d} K_k \dot{\phi}_{\bot {\cal D}} = {\p P_{\cal D} \over \p \phi_{\bot {\cal D}}}
\\&&
\label{eomscalars3}
\ddot{\zeta}_p + (d-1) H \dot{\zeta}_p + \sum_{k=1}^{{\cal D} - d} K_k \dot{\zeta}_p = {\p P_{\cal D} \over \p \zeta_p} \qquad \qquad (p = 1,...,\Delta)
\\&&
\label{eomscalars4}
\ddot{\varphi}_i + (d-1) H \dot{\varphi}_i + \sum_{k=1}^{{\cal D} - d} K_k \dot{\varphi}_i = {\p P_{\cal D} \over \p \varphi_i} \qquad \qquad \!(i = 1,...,n-1).
\eea


\subsection{Reduced equations of motion for  phases I and II}
\label{eq d}

Here, we apply the above results to the background of Sects \ref{Higgs} and \ref{plateau}. We have $n=1$ internal direction that breaks supersymmetry and $D=8-A$ ``spectator'' directions ($A=0$ for the toroidal models and $A=4$ for the orbifold ones). We split $T^D\equiv T^{d-1}\times \big(S^1(R_d)\times T^{D-d}\big)$, where $T^{d-1}$ is very large, the radii of $T^{D-d}$ are small and frozen, and we are interested in the regime where $R_d$ and $1/R_d$ are smaller than $\inf_{i\in I_b}R_i$. We thus choose $\Delta=9-A-d$ and find an effective space-time dimension $\D=d$. Beside the temperature $T_d$, the independent fields are the scale factor $a$ and the scalars $\Phi_d$, $\phi_{\bot d}$ and $\zeta_1=\ln R_d$, whose expressions follow from Eq. (\ref{orthofi}), with $\alpha = \sqrt{(d-1) / (d - 2)}$,
\bea
\Phi_{d} &\equiv& {2 \over \sqrt{({d} - 2)({d} - 1)}} \dil {}_{d} - \sqrt{{d} - 2 \over {d} -1} \ln R_9
\no\\
\phi_{\bot d} &\equiv& {2 \over \sqrt{{d} -1}} \dil {}_{{d}} + {1 \over \sqrt{{d} -1}} \ln R_9
\\
\zeta_1 &\equiv& \ln R_{d}.\no
\eea
To simplify the notations, we will denote the fields as $T,a,\Phi,\phi_\bot,\zeta$ and the thermal energy density and pressure as $\rho$, $P$.

The two Einstein equations (\ref{eomeinstein}) and (\ref{eomeinstein2}) simplify and can be replaced by the Friedmann equation and an equation expressing the conservation of energy,
\be
\label{eq1'}
{1\over 2}(d-2)(d-1)H^2 = {1\over 2}\left( \dot\Phi^2+\dot\phi_\bot^2+\dot\zeta^2\right) + \rho \, ,
\ee
\be
\label{eq2''}
\dot \rho +(d-1)H(\rho+P)+\dot \Phi \,  {\partial P\over \partial \Phi}+\dot \phi_\bot\, {\partial P\over \partial \phi_\bot}+\dot \zeta\, {\partial P\over \partial \zeta}  =0,
\ee
where the sources $P, \rho$ satisfy (see Sect. \ref{form1})
\be
\label{rp}
P=T^d\, p(z,\eta,\zeta),\; \rho=T\, {\partial P\over \partial T}-P\quad \Longrightarrow\quad \rho=T^d\, r(z,\eta,\zeta)\quad\with\quad r=(d-1)p-p_z,
\ee
where $\eta=\ln R_9$ and $e^z$ is the ratio of temperature to supersymmetry breaking \ie $z = \sqrt{d-1 \over d-2} \Phi - \ln(2 \pi T)$.\footnote{In this section, it is understood that partial derivatives of $p$ are with respect to $z$, $\eta$ or $\zeta$ with the remaining variables held constant.}
The scalar field equations (\ref{eomscalars})--(\ref{eomscalars3}) reduce to
\ba
\label{eq3}
\ddot \Phi+(d-1)H\dot \Phi\!\!\! &=&\!\!\! {\partial P\over \partial \Phi}=T^d\,  \left(\sqrt{d-1\over d-2}\, p_z-\sqrt{d-2\over d-1}\, p_\eta\right) ,\\
\label{eq4}
\ddot \phi_\bot+(d-1)H\dot \phi_\bot\!\!\! &=&\!\!\! {\partial P\over \partial \phi_\bot}={T^d\over \sqrt{d-1}}\, p_\eta\, ,\\
\label{eq5}
\ddot \zeta+(d-1)H\dot \zeta\!\!\! &=&\!\!\! {\partial P\over \partial \zeta}=T^d\, p_\zeta\, .
\ea
Note that Eq. (\ref{eq2''}) can easily be integrated. Writing $\dis \dot P = \dot \Phi \,  {\partial P\over \partial \Phi}+ \dot \phi_\bot\, {\partial P\over \partial \phi_\bot}+\dot \zeta\, {\partial P\over \partial \zeta} + \dot T \, {\partial P \over \partial T}$, one derives from Eq. (\ref{eq2''}) and the relation between $\rho$ and $P$ in Eq. (\ref{rp}),
\be
\label{aTcst}
\dot \rho + \dot P +(d-1)H(\rho+P) = {\dot T \over T} \, (\rho + P) \quad \Longrightarrow\quad (aT)^{d-1}\big(r(z,\eta,\zeta)+p(z,\eta,\zeta)\big)=\rm cst.
\ee

It is useful to parameterize our functions in terms of $\ln a$ and thereby replace time-derivatives by $(\ln a)$-derivatives, in which case we have
\be
\dot f=H \, {df\over d\ln a} := H\, \o f\, ,
\ee
for any function $f$.
Critical solutions do not have constant $\Phi$ but rather constant $z$ and so it is relevant to change variables from $\Phi$ to $z$ in Eq. (\ref{eq3}). Physically, this corresponds to the fact that for stable solutions the ratio of the supersymmetry breaking scale to the temperature scale must be a constant.  In order to proceed, we first express the derivative of the energy density $\rho$ in $T$, $z$, $\eta$ and $\zeta$ variables as $\dot \rho=T^dH(d\, r\o T/T+r_z\o z+r_\eta\o\eta +r_\zeta \o \zeta)$ and note from the definition of $z$ that $\o T/T= \sqrt{(d-1)/(d-2)}\o \Phi-\o z$.  Using these two expressions, Eq. (\ref{eq2''}) can be reexpressed as
\be
\label{oPhi}
\o \Phi=\A_{(z)}\o z+\A_{(\phi_\bot)}\o \phi_\bot+\A_{(\zeta)}\o \zeta +\B\, ,
\ee
where
\be
\label{AB}
\begin{array}{ll}
\displaystyle \A_{(z)}(z,\eta,\zeta) ={d\, r-r_z\over \E}\; , &\displaystyle\A_{(\phi_\bot)}(z,\eta,\zeta)=-{1\over \sqrt{d-1}}\, {r_\eta+p_\eta \over \E}\, ,\\\\
\displaystyle \A_{(\zeta)}(z,\eta,\zeta)=-{r_\zeta+p_\zeta \over \E}\; , &\displaystyle\B(z,\eta,\zeta)=-(d-1)\, {r+p \over \E}\, ,
\end{array}
\ee
and
\be
\E=\sqrt{d-1\over d-2}\, (d \, r+p_z)-\sqrt{d-2\over d-1}\, (r_\eta+p_\eta)\, .
\ee
Eq. (\ref{oPhi}) can now be used to eliminate $\Phi$ from the equations.  We first use it to write the Friedmann Eq. (\ref{eq1'}) in the form,
\be
\label{h}
H^2=h \, T^d\qquad \where \qquad h(z,\eta,\zeta;\o z,\o \phi_\bot,\o \zeta)={r\over {1\over 2}(d-2)(d-1)-\K}
\ee
and
\be
\K={1\over 2}\left[\left(\A_{(z)}\o z+\A_{(\phi_\bot)}\o \phi_\bot+\A_{(\zeta)}\o \zeta +\B\right)^2+\o \phi_\bot \!\!\!{}^2+\o\zeta {}^2\right]\, .
\ee
Next, noting that $\ddot\Phi=\dot H\o \Phi+H^2\oo\Phi$, one can express $\dot H$ in terms of $H, \rho, P$ using Einstein's equations and bring (\ref{eq3}) into the form,
\be
\label{eq3o}
\begin{array}{l}
h\left[ \A_{(z)}\oo z+ \A_{(\phi_\bot)}\oo \phi_\bot +\A_{(\zeta)}\oo \zeta +(\o z,\o \phi_\bot,\o\zeta)\; \C
\right.\!\left(\!\!\!\begin{array}{l}\o z\\ \o\phi_\bot\\ \o \zeta\end{array}\!\!\!\!\!\right)\left.\phantom{\o\zeta}\!\!\!\!\right]\\
~~~~~+\left[h\left(\B_z-\sqrt{d-2\over d-1}\, \left(\A_{(z)}\B\right)_\eta\right)+{1\over d-2}\, (r-p)\A_{(z)}\right]\o z\\
~~~~~+\left[h\left({1\over \sqrt{d-1}}\, \B_\eta-\sqrt{d-2\over d-1}\, \left(\A_{(\phi_\bot)}\B\right)_\eta\right)+{1\over d-2}\, (r-p)\A_{(\phi_\bot)}\right]\o \phi_\bot\\
~~~~~+\left[h\left(\B_\zeta-\sqrt{d-2\over d-1}\, \left(\A_{(\zeta)}\B\right)_\eta\right)+{1\over d-2}\, (r-p)\A_{(\zeta)}\right]\o \zeta+V_z=0\, ,
\end{array}
\ee
where the matrix $\C$ is
\be
\label{C}
\left(
\begin{array}{lll}
\A_{(z)z}-\sqrt{d-2\over d-1}\A_{(z)}\A_{(z)\eta}& \A_{(z)\eta}\left({1\over \sqrt{d-1}}-\sqrt{d-2\over d-1}\A_{(\phi_\bot)}\right)&\A_{(z)\zeta}-\sqrt{d-2\over d-1}\A_{(\zeta)}\A_{(z)\eta} \\
\A_{(\phi_\bot)z}-\sqrt{d-2\over d-1}\A_{(z)}\A_{(\phi_\bot)\eta}& \A_{(\phi_\bot)\eta}\left({1\over \sqrt{d-1}}-\sqrt{d-2\over d-1}\A_{(\phi_\bot)}\right) &\A_{(\phi_\bot)\zeta}-\sqrt{d-2\over d-1}\A_{(\zeta)}\A_{(\phi_\bot)\eta} \\
\A_{(\zeta)z}-\sqrt{d-2\over d-1}\A_{(z)}\A_{(\zeta)\eta}& \A_{(\zeta)\eta}\left({1\over \sqrt{d-1}}-\sqrt{d-2\over d-1}\A_{(\phi_\bot)}\right) &\A_{(\zeta)\zeta}-\sqrt{d-2\over d-1}\A_{(\zeta)}\A_{(\zeta)\eta}
\end{array}
\right)
\ee
and we have introduced $V(z,\eta,\zeta;\o z,\o \phi_\bot,\o \zeta)$, whose derivative with respect to $z$ is
\be
\label{V_z}
V_z=-\sqrt{d-1\over d-2}\, p_z+\sqrt{d-2\over d-1}\, p_\eta+{1\over d-2}\, (r-p)\, \B-\sqrt{d-1\over d-2}\, h\, \B_\eta\, \B\, .
\ee
Similarly, the equations (\ref{eq4}), (\ref{eq5}) for $\phi_\bot$ and $\zeta$ become,
\ba
\label{eq4o}
&&h\,\oo\phi_\bot+{1\over d-2}\, (r-p)\, \o\phi_\bot-{1\over \sqrt{d-1}}\, p_\eta=0\, ,\\
\label{eq5o}
&&h\,\oo\zeta+{1\over d-2}\, (r-p)\, \o\zeta- p_\zeta=0\, .
\ea

Finally, note that for $\zeta = 0$, $V_z$ defined in (\ref{V_z}) does not depend on velocities anymore,
\be
\label{simpvz}
\left. V_z\right\abs_{\zeta=0} = \sqrt{d-2 \over d-1} \bigg( r(z,\eta,0) - d \, p(z,\eta,0) \bigg).
\ee


\subsection{Reduced equations of motion for phase III}
\label{eq d+1}

We want to write the fields and equations of motions in the regime of Sect. \ref{spatial}. Again, $n=1$ and $D=8-A$. We split $T^D$ as $\big(T^{d-1}\times S^1(R_d)\big)\times T^{D-d}$, where $T^{d-1}$ is very large, the radii of $T^{D-d}$ are small and frozen, and we are interested in the regime where $R_d$ (or its inverse) is large. We thus take $\Delta=8-A-d$, which implies an effective space-time dimension $\D=d+1$. The independent fields are the temperature $T_{d+1}$, the scale factor associated to the torus $T^{d-1}$, the scale factor $b_1$ of the spatial circle $S^1(R_d)$ and $\Phi_{d+1}$, $\phi_{\bot d+1}$. The definitions of the scalars is derived from Eq. (\ref{orthofi}), with $\alpha = \sqrt{d / (d - 1)}$,
\bea
\Phi_{d+1} &\equiv& {2 \over \sqrt{d({d} - 1)}} \dil {}_{d+1} - \sqrt{{d} - 1 \over {d} } \ln R_9
\no\\
\phi_{\bot d+1} &\equiv& {2 \over \sqrt{{d} }} \dil {}_{{d+1}} + {1 \over \sqrt{{d} }} \ln R_9.
\eea
We introduce simpler notations for the fields in $d+1$ dimensions, $T',a',b,\Phi',\phi'_\bot$, and for the thermal energy density and pressure, $\rho', P'$.

There are three Einstein equations (\ref{eomeinstein})--(\ref{eomeinstein3}) which simplify as,
\ba
\label{eq'1}
-(d-1)(\dot H'+H^{\prime 2})-(\dot K+K^2)\!\!\!&=&\!\!\!\dot\Phi^{\prime 2}+\dot\phi_\bot^{\prime 2}+{1\over d-1}\left((d-2)\rho'+d \, P'+b\, {\partial P'\over \partial b}\right), \\
\label{eq'2}
\dot H'+(d-1)H^{\prime 2}+H'K\!\!&=&\!\!{1\over d-1}\left( \rho'-P'-b\, {\partial P'\over \partial b}\right) ,\\
\label{eq'3}
\dot K+(d-1)H'K+K^2\!\!&=&\!\!{1\over d-1}\left( \rho'-P'+(d-2)\, b\, {\partial P'\over \partial b}\right) ,
\ea
where $\dis H'={\dot a'\over a'}$ and $\dis K={\dot b\over b}$. The dependencies of the thermal sources $P'$ and $\rho'$ are as follows (see Sect. \ref{spatial}),
\be
P'=T^{\prime d+1}p'(z,\eta,\abs\zeta\abs),\; \rho'=T'\, {\partial P'\over \partial T'}-P'\quad \Longrightarrow\quad \rho'=T^{\prime d+1}\, r'(z,\eta,\abs\zeta\abs)\quad\with\quad r'=d\, p'-p'_z,
\ee
where $\eta=\ln R_9$, $\zeta=\ln R_d$ is related to the definition of $b$ via $\abs\zeta\abs=\ln b+{1\over \sqrt{d(d-1)}}\Phi'+{1\over \sqrt{d}}\phi'_\bot$ and $e^z$ is the ratio of temperature to supersymmetry breaking \ie $z = \sqrt{d \over d-1} \Phi' - \ln(2 \pi T')$.\footnote{In this section, it is understood that partial derivatives of $p'$ with respect to $z$, $\eta$ or $\abs\zeta\abs$ are with the two other variables held constant.}
The scalar equations of motion given in (\ref{eomscalars}) and (\ref{eomscalars2}) become in the present case,
\ba
\label{eq'4}
\ddot \Phi'+\left((d-1)H'+K\phantom{\dot \Phi}\!\!\!\!\right)\dot \Phi' \!\!\! &=&\!\!\! {\partial P'\over \partial \Phi'}\no\\&=&T^{\prime d+1}\!\!  \left(\!\sqrt{d\over d-1}\, p'_z-\sqrt{d-1\over d}\, p'_\eta+{1\over \sqrt{(d-1)d}}\, p'_{\abs\zeta\abs}\!\right),\\
\label{eq'5}
\ddot \phi'_\bot+\left((d-1)H'+K\phantom{\dot \Phi}\!\!\!\!\right)\dot \phi'_\bot \!\!\! &=&\!\!\! {\partial P'\over \partial \phi'_\bot}\no\\&=&{T^{\prime d+1}\over \sqrt{d}}\, \left(p'_\eta+p'_{\abs\zeta\abs}\right)\, .
\ea
Again, it is convenient to derive from Eqs. (\ref{eq'1})--(\ref{eq'5}) the Friedmann equation and the conservation of the energy-momentum tensor,
\be
\label{eq'1'}
{1\over 2}(d-1)\left((d-2)H^{\prime 2}+2HK\phantom{\dot \Phi}\!\!\!\!\right) ={1\over 2}\left( \dot\Phi^{\prime 2}+\dot\phi_\bot^{\prime 2}\right) + \rho'  ,
\ee
\be
\label{eq'2'''}
\dot \rho' +\left((d-1)H'+K\phantom{\dot \Phi}\!\!\!\!\right)\left(\rho'+P'\right)+\dot \Phi' \,  {\partial P'\over \partial \Phi'}+\dot \phi'_\bot\, {\partial P'\over \partial \phi'_\bot}+\dot b\, {\partial P'\over \partial b}  =0.
\ee
To obtain a useful form for the remaining independent equations, we first define
\be
e^\xi:= {b\over a'}\qquad\Longrightarrow\qquad K\equiv H'+\dot\xi\, ,
\ee
and then subtract Eq. (\ref{eq'2}) from (\ref{eq'3}) to obtain,
\be
 \label{eq'3''}
\ddot \xi+\left((d-1)H'+K\phantom{\dot \Phi}\!\!\!\!\right)\dot \xi = b\, {\partial P'\over \partial  b}=T^{\prime d+1}\, p'_{\abs\zeta\abs}\, .
\ee
Eq. (\ref{eq'2'''}) can be integrated by writing $\dis \dot P{}' = \dot \Phi{}' \,  {\partial P'\over \partial \Phi'}+ \dot \phi{}'_\bot\, {\partial P'\over \partial \phi'_\bot}+\dot b\, {\partial P'\over \partial b} + \dot T{}' \, {\partial P' \over \partial T'}$ to obtain,
\be
\label{aTxicst}
\dot \rho{}' + \dot P{}' +(d\, H'+\dot\xi)(\rho'+P') = {\dot T{}' \over T'} \, (\rho' + P') \quad \Longrightarrow\quad (a'T')^{d}\, e^\xi\, \big(r'(z,\eta,\abs\zeta\abs)+p'(z,\eta,\abs\zeta\abs)\big)=\rm cst.
\ee

As in appendix \ref{eq d}, we introduce  $(\ln a')$-derivatives, where for any function $f$,
\be
\dot f=H' \, {df\over d\ln a'} := H'\, \o f\, .
\ee
Proceeding in an analogous manner as the derivation of (\ref{oPhi}), Eq. (\ref{eq'2'''}) can also be rewritten as
\be
\label{oPhi'}
\o \Phi{}'=\A'_{(z)}\o z+\A'_{(\phi'_\bot)}\o \phi_\bot{\!\!\!}'+\A'_{(\xi)}\o \xi +\B'\, ,
\ee
where
\be
\label{A'B'}
\begin{array}{ll}
\displaystyle\A'_{(z)}(z,\eta,\abs\zeta\abs) ={(d+1)\, r'-r'_z\over \E'}\; , &\displaystyle\A'_{(\phi'_\bot)}(z,\eta,\abs\zeta\abs)=-{1\over \sqrt{d}}\, {r'_\eta+p'_\eta \over \E'}\, ,\\\\
\displaystyle\A'_{(\xi)}(z,\eta,\abs\zeta\abs)=-{r'+p'+r'_{\abs\zeta\abs}+p'_{\abs\zeta\abs} \over\E'}\; , &\displaystyle\B'(z,\eta,\abs\zeta\abs)=-{d\, (r'+p')+r'_{\abs\zeta\abs}+p'_{\abs\zeta\abs} \over \E'}\, ,
\end{array}
\ee
and
\be
\E'= \sqrt{d\over d-1}\, ((d+1) \, r'+p'_z)-\sqrt{d-1\over d}\, (r'_\eta+p'_\eta)+{1\over \sqrt{(d-1)d}}\, (r'_{\abs\zeta\abs}+p'_{\abs\zeta\abs})\, .
\ee
Eq. (\ref{oPhi'}) can be used to recast the Friedmann equation (\ref{eq'1'}) in the form,
\be
\label{h'}
H^{\prime 2}=h' \, T^{\prime d+1}\qquad \where \qquad h'(z,\eta,\abs\zeta\abs;\o z,\o \phi_\bot,\o \xi)={r'\over {1\over 2}(d-1)d-\K'},
\ee
where
\be
\K'={1\over 2}\left[\left(\A'_{(z)}\o z+\A'_{(\phi'_\bot)}\o \phi_\bot\!\!\!{}^{\prime 2}+\A'_{(\xi)}\o \xi +\B'\right)^2+\o \phi_\bot \!\!\!{}^{\prime 2}-(d-1)\, \o\xi {}^2\right].
\ee
For the scalar $\Phi'$, its equation becomes
\be
\label{eq'4o}
\begin{array}{l}
h'\left[ \A'_{(z)}\oo z+ \A'_{(\phi'_\bot)}\oo \phi_\bot {\!\!\!}'\,\, +\A'_{(\xi)}\oo \xi +(\o z,\o \phi_\bot {\!\!\!}'\,\,,\o\xi)\; \C'
\right.\!\left(\!\!\!\begin{array}{l}\o z\\ \o\phi_\bot {\!\!\!}'\,\,\\ \o \xi\end{array}\!\!\!\!\!\right)\left.\phantom{\o\zeta}\!\!\!\!\right]\\
+\left[h'\left(\B'_z-\sqrt{d-1\over d} \left(\A'_{(z)}\B'\right)_\eta+{1\over\sqrt{(d-1)d}}\left(\A'_{(z)}\B'\right)_{\abs \zeta\abs}+\A'_{(z)\abs\zeta\abs}\right)+{(r'-p'-p'_{\abs\zeta\abs})\over d-1}\A'_{(z)}\right]\o z\\
+\left[h'\left({1\over \sqrt{d}}\left(\B'_\eta+\B'_{\abs\zeta\abs}\right)-\sqrt{d-1\over d}\left(\A'_{(\phi'_\bot)}\B'\right)_\eta+{1\over\sqrt{(d-1)d}}\left(\A'_{(\phi'_\bot)}\B'\right)_{\abs \zeta\abs}+\A'_{(\phi'_\bot)\abs\zeta\abs}\right)+{(r'-p'-p'_{\abs\zeta\abs})\over d-1}\A'_{(\phi'_\bot)}\right]\o \phi_\bot{\!\!\!}'\\
+\left[h'\left(\B'_{\abs\zeta\abs}-\sqrt{d-1\over d}\left(\A'_{(\xi)}\B'\right)_\eta+{1\over\sqrt{(d-1)d}}\left(\A'_{(\xi)}\B'\right)_{\abs \zeta\abs}+\A'_{(\xi)\abs\zeta\abs}\right)+{(r'-p'-p'_{\abs\zeta\abs})\over d-1}\A'_{(\xi)}\right]\o \xi+V'_{z}=0,
\end{array}
\ee
where the components of the matrix $\C'$ are collected below, in Eqs (\ref{Ci1})--(\ref{Ci3}) and $V'(z,\eta,\abs \zeta\abs;\o z,\o \phi_\bot,\o \zeta)$ is defined by
\be
\label{V'_z}
\begin{array}{ll}
 V'_z=&\!\!\! \displaystyle-\sqrt{d\over d-1}\, p'_z+\sqrt{d-1\over d}\, p'_\eta-{1\over \sqrt{(d-1)d}}\, p'_{\abs\zeta\abs}\\&\!\!\!\displaystyle+{1\over d-1} \left(r'-p'-p'_{\abs\zeta\abs}\right) \B'+h'\left[  \B'_{\abs\zeta\abs}\left({1\over\sqrt{(d-1)d}}\, \B'+1\right)-\sqrt{d-1\over d}\, \B'_\eta\, \B'\right] .
\end{array}
\ee
The remaining equations (\ref{eq'5}) for $\phi'_\bot$ and (\ref{eq'3''}) for $\xi$ take the form
\ba
\label{eq'5o}
&&h'\, \oo\phi_\bot{\!\!\!}'\,\,+{1\over d-1}\, \left(r'-p'-p'_{\abs\zeta\abs}\right)\, \o\phi_\bot{\!\!\!}'\,\,-{1\over \sqrt{d}}\left(p'_\eta+p'_{\abs\zeta\abs}\right)=0\, ,\\
\label{eq'3''o}
&&h'\, \oo\xi+{1\over d-1}\, \left(r'-p'-p'_{\abs\zeta\abs}\right)\, \o\xi- p'_{\abs\zeta\abs}=0\, .
\ea
For completeness, we collect all entries of the matrix $\C'$ that appears in Eq. (\ref{eq'4o}):
\be
\label{Ci1}
\begin{array}{lll}
\C'_{zz}\!\!\! &=&\!\!\!\displaystyle \A'_{(z)z}-\A'_{(z)}\left(\sqrt{d-1\over d}\A'_{(z)\eta}-{1\over \sqrt{(d-1)d}}\A'_{(z)\abs\zeta\abs}\right)\\
\C'_{\phi'_\bot z}\!\!\! &=&\!\!\!\displaystyle \A'_{(\phi'_\bot)z}-\A'_{(z)}\left(\sqrt{d-1\over d}\A'_{(\phi'_\bot)\eta}-{1\over\sqrt{d-1)d}}\A'_{(\phi'_\bot)\abs\zeta\abs}\right)\\
\C'_{\xi z}\!\!\! &=&\!\!\!\displaystyle \A'_{(\xi)z}-\A'_{(z)}\left(\sqrt{d-1\over d}\A'_{(\xi)\eta}-{1\over\sqrt{(d-1)d}}\A'_{(\xi)\abs\zeta\abs}\right)
\end{array}
\ee
\be
\label{Ci2}
\begin{array}{lll}
\C'_{z\phi'_\bot}\!\!\! &=&\!\!\!\displaystyle {\A'_{(z)\eta}+\A'_{(z)\abs\zeta\abs}\over\sqrt{d}}-\A'_{(\phi'_\bot)}\left(\sqrt{d-1\over d} \A'_{(z)\eta}-{1\over\sqrt{(d-1)d}} \A'_{(z)\abs\zeta\abs}\right)\\
\C'_{\phi'_\bot \phi'_\bot}\!\!\! &=&\!\!\!\displaystyle  {\A'_{(\phi'_\bot)\eta}+\A'_{(\phi'_\bot)\abs\zeta\abs}\over\sqrt{d}}-\A'_{(\phi'_\bot)}\left(\sqrt{d-1\over d}\A'_{(\phi'_\bot)\eta}-{1\over \sqrt{(d-1)d}}\A'_{(\phi'_\bot)\abs\zeta\abs}\right)\\
\C'_{\xi \phi'_\bot}\!\!\! &=&\!\!\!\displaystyle{\A'_{(\xi)\eta}+\A'_{(\xi)\abs\zeta\abs}\over\sqrt{d}}-\A'_{(\phi'_\bot)}\left(\sqrt{d-1\over d}\A'_{(\xi)\eta}-{1\over\sqrt{(d-1)d}}\A'_{(\xi)\abs\zeta\abs}\right)
\end{array}
\ee
\be
\label{Ci3}
\begin{array}{lll}
\C'_{z\xi}\!\!\! &=&\!\!\!\displaystyle \A'_{(z)\abs\zeta\abs}-\A'_{(\xi)}\left(\sqrt{d-1\over d}\A'_{(z)\eta}-{1\over \sqrt{(d-1)d}}\A'_{(z)\abs\zeta\abs}\right)\\
\C'_{\phi'_\bot \xi}\!\!\! &=&\!\!\!\displaystyle \A'_{(\phi'_\bot)\abs\zeta\abs}-\A'_{(\xi)}\left(\sqrt{d-1\over d}\A'_{(\phi'_\bot)\eta}-{1\over \sqrt{(d-1)d}}\A'_{(\phi'_\bot)\abs\zeta\abs}\right)\\
\C'_{\xi \xi}\!\!\! &=&\!\!\!\displaystyle\A'_{(\xi)\abs\zeta\abs}-\A'_{(\xi)}\left(\sqrt{d-1\over d}\A'_{(\xi)\eta}-{1\over \sqrt{(d-1)d}}\A'_{(\xi)\abs\zeta\abs}\right).
\end{array}
\ee



\end{document}